\documentclass[prd,showpacs,preprintnumbers,amsmath,amssymb,eqsecnum]{revtex4-1}

\usepackage{graphicx}
\usepackage{dcolumn}
\usepackage{bm}
\usepackage{subfigure}
\usepackage{color}
\usepackage{ulem}

\begin{document}

\preprint{RUP-15-5}
\preprint{RESCEU-4/15}

\title{Cosmological long-wavelength solutions and 
primordial black hole formation}

\author{$^{1}$Tomohiro Harada}%
 \email{harada@rikkyo.ac.jp}
\author{$^{2}$Chul-Moon Yoo}
\email{yoo@gravity.phys.nagoya-u.ac.jp}
\author{$^{3}$Tomohiro Nakama}
\email{nakama@resceu.s.u-tokyo.ac.jp}
\author{$^{1}$Yasutaka Koga}
\email{koga@rikkyo.ac.jp}
\affiliation{$^{1}$Department of Physics, Rikkyo University, Toshima, Tokyo 171-8501, Japan} 
\affiliation{$^{2}$ Gravity and Particle Cosmology Group, Division of Particle and Astrophysical Science, Graduate School of Science, Nagoya University, Furo-cho, Chikusa-ku, Nagoya 464-8602, Japan}
\affiliation{$^{3}$ Research Center for the Early Universe (RESCEU), Graduate School of Science, University of Tokyo, Bunkyo-ku, Tokyo 113-0033, Japan}
\date{\today}

\begin{abstract}
We construct cosmological long-wavelength solutions without symmetry 
in general gauge conditions which are compatible with 
the long-wavelength scheme. We then specify the relationship 
among the solutions in different time slicings.
Nonspherical long-wavelength solutions are particularly
important for primordial structure formation in the epoch of very 
soft equations of state. 
Applying this general framework to spherical symmetry, 
 we show the equivalence between 
long-wavelength solutions in the constant mean
curvature slicing with conformally flat spatial coordinates 
and asymptotic quasihomogeneous solutions in the 
comoving slicing with the comoving threading.
We derive the correspondence relation between these two  
solutions and compare the results
of numerical simulations of primordial black hole (PBH) formation 
in these two different approaches.
To discuss the PBH formation, 
it is convenient and conventional to use $\tilde{\delta}_{c}$, 
the value which the averaged density perturbation at threshold 
in the comoving slicing would take at horizon entry 
in the lowest-order long-wavelength expansion. 
We numerically find that within 
(approximately) compensated models,  
the sharper the transition from the overdense region 
to the Friedmann-Robertson-Walker universe is,
the larger the $\tilde{\delta}_{c}$ becomes.
We suggest that, for the equation of state $p=(\Gamma-1)\rho$, 
we can apply the analytic formulas for the minimum 
$\tilde{\delta}_{c, {\rm  min}}\simeq [3\Gamma/(3\Gamma+2)]
 \sin^{2}\left[\pi\sqrt{\Gamma-1}/(3\Gamma-2)\right]$
and the maximum 
$\tilde{\delta}_{c, {\rm max}}\simeq 3\Gamma/(3\Gamma+2)$.
As for the threshold peak value 
of the curvature variable $\psi_{0,c}$, 
we find that the sharper the transition is, 
the smaller the $\psi_{0,c}$ becomes.
We analytically explain this intriguing feature 
qualitatively with a compensated top-hat density model.
Using simplified models, we also analytically 
deduce an environmental effect 
that $\psi_{0,c}$ can be significantly larger (smaller) if 
the underlying density perturbation of much longer wavelength is positive (negative).
\end{abstract}


\maketitle

\clearpage

\tableofcontents

\newpage

\section{Introduction}
The proposal that black holes may have formed in the early Universe, 
which are called primordial black holes (PBHs),
 has recently been extensively studied 
in cosmology because they can convey unique information about 
the early Universe to us. 
This is because energy scales relevant to PBHs are generally 
very different from those of the observed cosmic microwave background 
anisotropy.
The possibility of PBHs has been indicated by Zeldovich and
Novikov~\cite{Zeldovich:1967} and 
Hawking~\cite{Hawking:1971ei}.
Since PBHs evaporate by emitting radiation and they also act as 
gravitational sources before they evaporate away, 
current observations constrain the abundance of PBHs
using the big bang nucleosynthesis, extragalactic photon background
and gravitational and astrophysical effects of nonevaporating 
PBHs~\cite{Carr:1975qj,Carr:2009jm}.
To convert the observed abundance of PBHs to the information of the early
Universe, the key issues are the production efficiency of PBHs 
from given cosmological perturbations and the time evolution of PBHs 
through mass accretion and Hawking evaporation~\cite{Zeldovich:1967,Carr:1975qj,Carr:1974nx}.

Here, we focus on the criterion of PBH formation.
Since the probability of PBH formation is expected to be
exponentially small, the abundance of PBHs for given perturbations is 
sensitive to the formation criterion. 
As a pioneer, Carr~\cite{Carr:1975qj} 
derived $\delta_{c}\simeq \Gamma-1$ for the equation of state
$p=(\Gamma-1)\rho$ almost in an order-of-magnitude estimate, where $\delta_{c}$ is 
the threshold value of the density perturbation at horizon
entry. This gives $\delta_{c}\simeq 1/3$ 
for a radiation fluid $\Gamma=4/3$. In fact, 
it turns out that it is not so straightforward to accurately 
determine the formation criterion of PBHs as 
we have to understand highly general 
relativistic nonlinear dynamics in the 
cosmological background.
Recently, numerical relativity has developed so much 
that the formation of PBHs can be simulated and 
the threshold of PBH formation can be
obtained~\cite{Nadezhin:1978,Novikov:1980,Niemeyer:1999ak,Shibata:1999zs,Musco:2004ak,Polnarev:2006aa,Musco:2008hv,Polnarev:2012bi,Nakama:2013ica,Musco:2012au,Nakama:2014fra}. 
Shibata and Sasaki~\cite{Shibata:1999zs} gave the threshold 
$\psi_{0,c}\simeq 1.4-1.8$, where $\psi_{0,c}$ is the threshold value of
the peak value $\psi_{0}$ of the curvature variable $\psi$ 
and this is 
consistent with ${\delta}_{c}\simeq 0.3-0.5$~\cite{Green:2004wb,Young:2014ana}.
Polnarev and Musco~\cite{Polnarev:2006aa} gave 
$\tilde{\delta}_{c}\simeq 0.45-0.66$ for a radiation fluid, 
depending on the perturbation profiles,
where $\tilde{\delta}_{c}$
is the threshold value of the averaged density perturbation $\tilde{\delta}$
at horizon entry in the comoving slicing in the lowest-order
long-wavelength expansion.
Musco and Miller~\cite{Musco:2012au} gave $\tilde{\delta}_{c}$
for different values of $\Gamma$.
Nakama {\it et al.}~\cite{Nakama:2013ica}
investigated the PBH formation threshold for a much wider class 
of initial curvature perturbation 
profiles characterized by five parameters.
Harada {\it et al.}~\cite{Harada:2013epa} derived an analytic 
formula
$\tilde{\delta}_{c}=[3\Gamma/(3\Gamma+2)]\sin^{2}
\left[\pi\sqrt{\Gamma-1}/(3\Gamma-2)\right]$
under a certain set of assumptions, which gives $\tilde{\delta}_{c}\simeq 0.4135$ 
for a radiation fluid.
Young {\it et al.}~\cite{Young:2014ana} suggested
that there is a significant environmental effect on $\psi_{0,c}$ but 
not on $\tilde{\delta}_{c}$. Very recently, 
Tada and Yokoyama~\cite{Tada:2015noa} and 
Young and Byrnes~\cite{Young:2015kda} have found 
that the PBH production rate is significantly biased 
if the statistics of the primordial curvature perturbation 
has small local-type non-Gaussianity.  

To discuss the PBH formation or any other primordial structure
formation, it is important to give cosmological perturbations which
can be generated in the early Universe. On the other hand, the 
initial data set which will result in PBH formation must nonlinearly  
deviate from the Friedmann-Robertson-Walker (FRW) universe 
even much before the horizon entry of the scale of the perturbations.
Such cosmological primordial perturbations are formulated by Lyth {\it
et al.}~\cite{Lyth:2004gb}.
There have been two approaches so far in which the appropriate initial data
of cosmological perturbations are prepared and the Einstein equations 
are numerically solved in spherical symmetry to follow the formation 
of PBHs. The one is initiated by Shibata and Sasaki~\cite{Shibata:1999zs} and the other is by Polnarev and Musco~\cite{Polnarev:2006aa}. Although these two approaches study 
the same problem, the direct comparison of their results has not been
done and for this reason the complete picture of PBH formation has yet 
been unclear.

In this paper, without assuming symmetry, we construct cosmological 
long-wavelength solutions which 
can deviate from the FRW spacetimes with an arbitrarily large
amplitude and are naturally generated in inflationary
cosmology. These solutions can be applied as the initial data 
for PBH formation and any other primordial structure formation 
in spherical symmetry or in nonspherical situations.
Nonspherical cosmological perturbations should be particularly important 
for PBH formation in the stage when the effective equation of state
is very soft~\cite{Khlopov:1980mg,Polnarev:1981,Polnarev:1982}. 
Such stages have recently been
discussed 
in some scenarios of inflationary cosmology~\cite{Suyama:2004mz,Suyama:2006sr,Alabidi:2012ex,Alabidi:2013lya}.

We explicitly show that the solutions in spherical symmetry 
reduce to the ones in Shibata and Sasaki~\cite{Shibata:1999zs} in some gauge and 
to the ones in Polnarev and Musco~\cite{Polnarev:2006aa} in another gauge. 
Thus, we 
derive the correspondence relation between these two approaches.
Based on this correspondence relation, we compare the numerical results 
obtained in these two approaches and find that the results are
consistent. We analyze three different measures of the amplitude of 
perturbations, $\tilde{\delta}$, $\psi_{0}$
and the maximum value of the compaction function ${\cal C}_{\rm max}$.
The threshold values of these three quantities depend on 
the initial density (or curvature) profiles.
We analyze how the threshold values depend on the profiles
and physically understand the dependence using simplified 
models. We also discuss an environmental or bias effect on 
$\psi_{0,c}$ 
when perturbations are superimposed on perturbations
of longer wavelength. 

This paper is organized as follows. 
In Sec. II,  we present the 3+1 formulation of the Einstein equations
and the cosmological conformal decomposition. In Sec. III, we review
the long-wavelength scheme. In Sec. IV, we present the long-wavelength solutions in different gauge conditions.
In Sec. V, we present the gauge conditions adopted in 
Shibata and Sasaki~\cite{Shibata:1999zs}
and Polnarev and Musco~\cite{Polnarev:2006aa}
and find the relation among quasilocal quantities in spherical
symmetry. In Sec. VI, we show the equivalence between the 
long-wavelength solutions obtained in the two approaches 
and derive the correspondence relation.
 In Sec. VII, we compare and interpret the numerical results obtained in the two
 approaches. In Sec. VIII, we discuss the environmental effect on
 $\psi_{0,c}$ with several simplified models.
We use the units in which $G=c=1$ and the signature convention $(-+++)$.

\section{Basic equations}

\subsection{3+1 formalism}
We here present the 3+1 formalism of the Einstein equations according to
Nakamura {\it et al.}~\cite{Nakamura:1987zz}.
The line element in four-dimensional spacetimes is written in the
following form:
\begin{eqnarray}
 ds^{2}=-\alpha^{2}dt^{2}+\gamma_{ij}(dx^{i}+\beta^{i}dt)(dx^{j}+\beta^{j}dt),
\end{eqnarray}
where $\alpha$, $\beta^{i}$ and $\gamma_{ij}$
are the lapse function, shift vector and spatial metric, respectively.
The Latin uppercase indices run over 1 to 3
and we drop and raise them by $\gamma_{ij}$ and
its inverse $\gamma^{ij}$, respectively, unless otherwise specified.
The spacetime metric $g_{\mu\nu}$ and its inverse $g^{\mu\nu}$ are given by 
\begin{eqnarray}
 g_{\mu\nu}=\left(\begin{array}{cc}
  -\alpha^{2}+\beta_{i}\beta^{i}& \beta_{i} \\
  \beta_{j} & \gamma_{ij} 
	      \end{array} 
\right) \quad\mbox{and} 
 g^{\mu\nu}=\left(\begin{array}{cc}
  -\displaystyle\frac{1}{\alpha^{2}} & \displaystyle\frac{\beta^{i}}{\alpha^{2}} \\
  \displaystyle\frac{\beta^{j}}{\alpha^{2}} & \gamma^{ij}-\displaystyle\frac{\beta^{i}\beta^{j}}{\alpha^{2}} 
	      \end{array}
\right),
\end{eqnarray} 
respectively, where the Greek indices run over 0 to 3. 
Because of the construction of the inverse matrix, we find 
$
 g=-\alpha^{2}\gamma,
$
where $g=\det(g_{\mu\nu})$ and $\gamma=\det (\gamma_{ij})$.
The covariant and contravariant components of the 
normal unit vector to the $t=$const hypersurface $\Sigma$ are given by 
\begin{equation}
 n_{\mu}=(-\alpha,0,0,0) \quad\mbox{and}\quad 
  n^{\mu}=\left(\frac{1}{\alpha},-\frac{\beta^{i}}{\alpha}\right),
\label{eq:n^mu}
\end{equation}
respectively.
The projection tensor to $\Sigma$ is defined as
$
 h_{\mu\nu}:=g_{\mu\nu}+n_{\mu}n_{\nu}.
$

The stress-energy tensor for the matter fields $T_{\mu\nu}$
is decomposed into the following form:
\begin{equation}
 T_{\mu\nu}=E n_{\mu}n_{\nu}+J_{\mu}n_{\nu}+J_{\nu}n_{\mu}
+S_{\alpha\beta}h_{\mu}^{\alpha}h_{\nu}^{\beta},
\end{equation}
where 
$ E :=T_{\mu\nu}n^{\mu}n^{\nu}$, 
$J_{\alpha}:=-T_{\mu\nu}h_{\alpha}^{\mu}n^{\nu}$
and 
$
 S_{\alpha\beta}:=T_{\mu\nu}h^{\mu}_{\alpha}h^{\nu}_{\beta}$.

The Einstein equation 
$
 G_{\mu\nu}=8\pi T_{\mu\nu}
$
can be written in the following set of equations. The Hamiltonian 
constraint $G^{\mu\nu}n_{\mu}n_{\nu}=8\pi T^{\mu\nu}n_{\mu}n_{\nu}$ 
and momentum constraint $G^{\mu\nu}n_{\mu}h_{\nu i}=8\pi
T^{\mu\nu}n_{\mu}h_{\nu i}$ reduce to 
\begin{equation}
 {\cal R}+K^{2}-K_{ij}K^{ij}=16\pi E
\label{eq:Hamiltonian_constraint}
\end{equation}
and 
\begin{equation}
 {\cal D}_{j}K^{j}_{i}-{\cal D}_{i}K=8\pi J_{i},
\label{eq:momentum_constraint}
\end{equation}
respectively, 
where ${\cal D}_{i}$ and ${\cal R}$ denote 
the covariant derivative and Ricci
scalar with respect to $\gamma_{ij}$, respectively, 
$K_{ij}$ is the extrinsic curvature of $\Sigma$ defined by 
\begin{equation}
 K_{ij}:=-h^{\mu}_{i}h^{\nu}_{j}n_{\mu;\nu}=-\frac{1}{2\alpha}(\gamma_{ij,t}
-{\cal  D}_{j}\beta_{i}-{\cal D}_{i}\beta_{j}),
\label{eq:extrinsic_curvature}
\end{equation}
and $K:=\gamma^{ij}K_{ij}$.
Note that the semicolon 
denotes the covariant derivative with respect to $g_{\mu\nu}$.

The evolution equations $G^{\mu\nu}h_{\mu i}h_{\nu j}=8\pi
T^{\mu\nu}h_{\mu i}h_{\nu j}$ are given by 
\begin{equation}
 K_{ij,t}=\alpha({\cal R}_{ij}+KK_{ij})-2\alpha K_{il}K^{l}_{j}-8\pi
  \alpha \left[S_{ij}+\frac{1}{2}\gamma_{ij}(E -S^{l}_{l})\right]
  -{\cal D}_{j}{\cal D}_{i}\alpha+({\cal D}_{j}\beta^{m})K_{mi}+({\cal
  D}_{i}\beta^{m})K_{mj}+\beta^{m}{\cal D}_{m}K_{ij},
\label{eq:evolution_equation}
\end{equation}
where ${\cal R}_{ij}$ is the Ricci tensor with respect to $\gamma_{ij}$
on $\Sigma$ with ${\cal R}=\gamma^{ij}{\cal R}_{ij}$.
Equation~(\ref{eq:extrinsic_curvature}) can be rewritten in the following form:
\begin{equation}
 \gamma_{ij,t}=-2\alpha K_{ij}+{\cal D}_{j}\beta_{i}+{\cal D}_{i}\beta_{j}.
\end{equation}

The conservation law 
$
 T^{\mu\nu}_{~~~~;\nu}=0
$
is decomposed into $T^{\mu\nu}_{~~~~;\nu}n_{\mu}=0$ and
$T^{\mu\nu}_{~~~~;\nu}h_{\mu m}=0$. 
Assuming a perfect fluid, 
\begin{equation}
 T_{\mu\nu}=(\rho +p)u_{\mu}u_{\nu}+p g_{\mu\nu},
\label{eq:perfect_fluid}
\end{equation} 
where $u^{\mu}$ is the four-velocity of the fluid element which is 
normalized as $u^{\mu}u_{\mu}=-1$,  
they reduce to
\begin{eqnarray}
(\sqrt{\gamma}E )_{,t}+(\sqrt{\gamma}E v^{l})_{,l}
=-[\sqrt{\gamma}p(v^{l}+\beta^{l})]_{,l}+\alpha\sqrt{\gamma}pK-\alpha_{,l}\sqrt{\gamma}J^{l}+\frac{\alpha\sqrt{\gamma}J^{l}J^{m}K_{lm}}{E
+p}
\label{eq:Energy_eq}
\end{eqnarray}
and 
\begin{equation}
 (\sqrt{\gamma}J_{m})_{,t}+(\sqrt{\gamma}J_{m}v^{l})_{,l}=-\alpha\sqrt{\gamma}p_{,m}
-\sqrt{\gamma}(p+E
)\alpha_{,m}+\frac{\alpha\sqrt{\gamma}\gamma_{kl,m}J^{k}J^{l}}{2(p+E
)}+\sqrt{\gamma}J_{l}\beta_{,m}^{l},
\label{eq:Euler_eq}
\end{equation}
respectively, 
where 
\begin{equation}
 v^{l}:=\frac{u^{l}}{u^{t}}
\label{eq:v^l}
\end{equation}
is the coordinate three-velocity of the fluid element. 
It is often useful to define the
``baryon'' or conserved number density $n$ and the conservation law 
$(n u^{\mu})_{;\mu}=0$ gives 
\begin{equation}
 (\sqrt{\gamma }\alpha u^{t}n)_{,t}+(\sqrt{\gamma}\alpha u^{t}n
  v^{l})_{,l}=0.
\label{eq:Baryon_number_conservation}
\end{equation} 
The energy conservation gives
\begin{equation}
 \frac{dn}{n}=\frac{d\rho}{\rho+p}.
\label{eq:Energy_conservation}
\end{equation}

The expressions for the four-velocity of the fluid element 
are rather complicated. In fact,  
Eq.~(\ref{eq:v^l}) implies (Lyth {\it et al.}~\cite{Lyth:2004gb})
\begin{eqnarray}
 u^{t}&=&\frac{1}{\sqrt{\alpha^{2}-(\beta_{k}+v_{k})(\beta^{k}+v^{k})}}, \\
 u^{i}&=&u^{t}v^{i}, \label{eq:u^i}\\
 u_{t}&=&-u^{t}[\alpha^{2}-\beta^{k}(\beta_{k}+v_{k})], \\
 u_{i}&=& u^{t}(v_{i}+\beta_{i}) \label{eq:u_i},
\end{eqnarray}
where $v_{i}=\gamma_{ij}v^{j}$. 
We can write down $E $, $J_{i}$ and $S_{ij}$ as
\begin{eqnarray}
 E &=&(\rho+p)w^{2}-p, 
\label{eq:E}\\
 J_{i}&=&\frac{w^{2}}{\alpha}(\rho+p)(v_{i}+\beta_{i}),
  \label{eq:J_i} 
\end{eqnarray}
and 
\begin{eqnarray}
 S_{ij}
=p\gamma_{ij}+\frac{J_{i}J_{j}}{E+p},
\label{eq:S_ij}
\end{eqnarray}
respectively, where 
\begin{equation}
w:=\alpha u^{t}=\left[1-\alpha^{-2}(\beta_{k}+v_{k})(\beta^{k}+v^{k})\right]^{-1/2}.
\label{eq:w}
\end{equation}

\subsection{Cosmological conformal decomposition
  \label{subsec:cosmological_conformal_decomposition}}
Here we briefly revisit the cosmological conformal decomposition
according to Shibata and Sasaki~\cite{Shibata:1999zs}.
See also Gourgoulhon~\cite{Gourgoulhon:2007ue}.
We decompose the spatial metric $\gamma_{ij}$ into the following form:
\begin{equation}
 \gamma_{ij}=\psi^{4}a^{2}(t)\tilde{\gamma}_{ij}.
\end{equation}
We choose $\tilde{\gamma}_{ij}$ so that 
$\tilde{\gamma}=\det(\tilde{\gamma}_{ij})$ is time independent 
and equal to $\eta=\det(\eta_{ij})$, where $\eta_{ij}$ is a
time-independent metric of the flat three-space.
The function $a(t)$ is the scale factor of the reference universe. 
The extrinsic curvature is decomposed as 
 \begin{equation}
 K_{ij}=A_{ij}+\frac{\gamma_{ij}}{3}K,
\label{eq:A_ij}
\end{equation}
where $A_{ij}$ is traceless by definition.
We also define $\tilde{A}_{ij}$ as follows:
\begin{equation}
 A^{ij}=\psi^{-4}a^{-2}\tilde{A}^{ij}\quad \mbox{or} \quad 
 A_{ij}=\psi^{4}a^{2}\tilde{A}_{ij}.
\label{eq:Atilde_ij}
\end{equation}
Thus, $\tilde{\gamma}^{ij}\tilde{A}_{ij}=0$ by definition.
We raise and drop the Latin lowercase 
indices $i,j,k,\ldots$ of tilded quantities 
by $\tilde{\gamma}^{ij}$ and $\tilde{\gamma}_{ij}$, respectively.
We define $\bar{\cal D}_{i}$ and $\tilde{\cal D}_{i}$ 
as the covariant derivatives with respect to $\eta_{ij}$
and $\tilde{\gamma}_{ij}$, respectively. 
We denote the Laplacians with respect to $\eta_{ij}$ and $\tilde{\gamma}_{ij}$
as $\bar{\Delta}:=\eta^{ij}\bar{\cal D}_{i}\bar{\cal D}_{j}$ and 
$\tilde{\Delta}:=\tilde{\gamma}^{ij}\tilde{\cal D}_{i}\tilde{\cal D}_{j}$, respectively.

Using this decomposition, we can rewrite ${\cal
R}_{ij}$ as follows:
\begin{equation}
 {\cal R}_{ij}=\tilde{{\cal R}}_{ij}+{\cal R}^{\psi}_{ij},
\end{equation}
where 
\begin{eqnarray}
 {\cal R}^{\psi}_{ij}&:=&-\frac{2}{\psi}\tilde{\cal D}_{i}\tilde{\cal D}_{j}\psi
-\frac{2}{\psi}\tilde{\gamma}_{ij}\tilde{\Delta}\psi
+\frac{6}{\psi^{2}}\tilde{\cal D}_{i}\psi \tilde{\cal D}_{j}\psi
-\frac{2}{\psi^{2}}\tilde{\gamma}_{ij}\tilde{\cal D}_{k}\psi
\tilde{\cal D}^{k}\psi, \\
 \tilde{{\cal R}}_{ij}&:=&\frac{1}{2}[-\bar{\Delta}\tilde{\gamma}_{ij}
  + \bar{\cal D}_{j} \bar{\cal D}^{k}\tilde{\gamma}_{ki}+\bar{\cal D}_{i}
  \bar{\cal D}^{k}\tilde{\gamma}_{kj}
+2 \bar{\cal D}_{k}(f^{kl}C_{lij})-2 C^{l}_{~~kj}C^{k}_{~~il}],
\label{eq:tildeRij} \\
 f^{kl}&:=&\tilde{\gamma}^{kl}-\eta^{kl}, \\
 C^{k}_{~~ij}&:=&\frac{1}{2}\tilde{\gamma}^{kl}(\bar{\cal D}_{i}\tilde{\gamma}_{jl}
+\bar{\cal D}_{j}\tilde{\gamma}_{il}-\bar{\cal D}_{l}\tilde{\gamma}_{ij}).
\end{eqnarray}
To derive Eq.~(\ref{eq:tildeRij}), we have used the relation
\begin{equation}
 \tilde{\gamma}^{ij}\bar{\cal D}_{k}\tilde{\gamma}_{ij}=\frac{1}{\tilde{\gamma}} \bar{\cal D}_{k}\tilde{\gamma}
 =\frac{1}{\eta} \bar{\cal D}_{k}\eta=
 \eta^{ij} \bar{\cal D}_{k}\eta_{ij}=0.
\end{equation}
The following relations will be useful:
\begin{eqnarray}
&& {\cal R}^{\psi}:=\gamma^{ij}{\cal R}^{\psi}_{ij}
=-\frac{8}{\psi^{5}a^{2}}\tilde{\Delta}\psi, \\
&&{\cal R}^{\psi}_{ij}-\frac{1}{3}\gamma_{ij}{\cal R}^{\psi}
=-\frac{2}{\psi}\left[\tilde{\cal D}_{i}\tilde{\cal D}_{j}\psi-\frac{1}{3}\tilde{\gamma}_{ij}\tilde{\Delta}\psi\right]+\frac{6}{\psi^{2}}\left[\tilde{\cal D}_{i}\psi\tilde{\cal D}_{j}\psi-\frac{1}{3}\tilde{\gamma}_{ij}\tilde{\cal D}^{k}\psi\tilde{\cal D}_{k}\psi\right].
\end{eqnarray}

The Hamiltonian and momentum constraints are given by 
\begin{eqnarray}
 {\cal R}_{k}^{k}-\tilde{A}_{ij}\tilde{A}^{ij}+\frac{2}{3}K^{2}=16\pi E, 
\label{eq:SS99_2.20}
\end{eqnarray}
and
\begin{equation}
 {\cal D}_{j}\tilde{A}^{j}_{i}-\frac{2}{3}{\cal D}_{i}K=8\pi J_{i},
\label{eq:SS99_2.21} 
\end{equation}
respectively.
These can be transformed to the following form:
\begin{eqnarray}
&& \tilde{\Delta}\psi=\frac{\tilde{\cal R}_{k}^{k}}{8}\psi-2\pi
  \psi^{5}a^{2}E
-\frac{\psi^{5}a^{2}}{8}\left(\tilde{A}_{ij}\tilde{A}^{ij}-\frac{2}{3}K^{2}\right), \label{eq:laplacianpsi}\\
&&
 \tilde{\cal D}^{j}(\psi^{6}\tilde{A}_{ij})-\frac{2}{3}\psi^{6}\tilde{\cal D}_{i}K=8\pi
 J_{i}\psi^{6}.
\label{eq:SS99_2.21_variant}
\end{eqnarray}
The evolution equations become
\begin{eqnarray}
 (\partial_{t}-{\cal L}_{\beta})\tilde{A}_{ij}&=&\frac{1}{a^{2}\psi^{4}}
\left[\alpha\left({\cal R}_{ij}-\frac{\gamma_{ij}}{3}{\cal R}\right)
-\left(D_{i}D_{j}\alpha-\frac{\gamma_{ij}}{3}D_{k}D^{k}\alpha\right)\right]
\nonumber \\
&& +\alpha(K\tilde{A}_{ij}-2\tilde{A}_{ik}\tilde{A}_{j}^{k})
-\frac{2}{3}(\bar{\cal D}_{k}\beta^{k})\tilde{A}_{ij}
-\frac{8\pi\alpha}{a^{2}\psi^{4}}\left(S_{ij}-\frac{\gamma_{ij}}{3}S_{k}^{k}\right), 
\label{eq:SS99_2.12}
\\
(\partial_{t}-{\cal L}_{\beta})\psi &=&
 -\frac{\dot{a}}{2a}\psi+\frac{\psi}{6}(-\alpha K+\bar{\cal D}_{k}\beta^{k}), \label{eq:SS99_2.13} \\
(\partial_{t}-{\cal L}_{\beta})K &=&
 \alpha\left(\tilde{A}_{ij}\tilde{A}^{ij}+\frac{1}{3}K^{2}\right)-D_{k}D^{k}\alpha
 +4\pi\alpha (E+S_{k}^{k}), 
\label{eq:SS99_2.14}
\end{eqnarray}
where ${\cal L}_{\beta}$ is the Lie derivative along $\beta^{i}$ and
acts on a scalar field $ f $ and a tensor field $f_{ij}$ as follows:
\begin{eqnarray}
 {\cal L}_{\beta}f&=&\beta^{k}f_{,k}, \\
 {\cal L}_{\beta}f_{ij}&=& \beta^{k}f_{ij,k}+\beta^{k}_{,i}f_{kj}
+\beta^{k}_{,j}f_{ki}.
\end{eqnarray}
The definition of the extrinsic curvature yields
\begin{equation}
 (\partial_{t}-{\cal L}_{\beta})\tilde{\gamma}_{ij}=-2\alpha
  \tilde{A}_{ij}-\frac{2}{3}\tilde{\gamma}_{ij}\bar{\cal D}_{k}\beta^{k}.
\label{eq:SS99_2.11}
\end{equation}

The hydrodynamical equations (\ref{eq:Energy_eq}),
(\ref{eq:Euler_eq}) and 
(\ref{eq:Baryon_number_conservation})
are, respectively, written in the form
\begin{eqnarray}
&&\left[\psi^{6}a^{3}\left\{(\rho+p)w^{2}-p\right\}\right]_{,t}
+\frac{1}{\sqrt{\eta}}\left[\sqrt{\eta}
\psi^{6}a^{3}\left\{(\rho+p)w^{2}-p\right\}v^{l}\right]_{,l}
\nonumber \\
&=&-\frac{1}{\sqrt{\eta}}\left[\sqrt{\eta}\psi^{6}a^{3}p(v^{l}+\beta^{l})\right]_{,l}
+\alpha\psi^{6}a^{3}pK
-\alpha^{-1}\alpha_{,l}\psi^{6}a^{3}w^{2}(\rho+p)(v^{l}+\beta^{l})
\nonumber \\
&&
 +\alpha^{-1}\psi^{10}a^{5}w^{2}(\rho+p)(v^{l}+\beta^{l})(v^{m}+\beta^{m})\left(\tilde{A}_{lm}+\frac{\tilde{\gamma}_{lm}}{3}K\right), 
\label{eq:Energy_eq_decomposition}\\
&&(w\psi^{6}a^{3}(\rho+p)u_{j})_{,t}+\frac{1}{\sqrt{\eta}}(\sqrt{\eta}
w\psi^{6}a^{3}(\rho+p)v^{k}u_{j})_{,k} \nonumber \\
&=&-\alpha\psi^{6}a^{3}p_{,j}+w\psi^{6}a^{3}(\rho+p)
\left(-w\alpha_{,j}+u_{k}\beta^{k}_{,j}-\frac{u_{k}u_{l}}{2u^{t}}\gamma^{kl}_{,j}\right),
\label{eq:SS99_2.9}
\end{eqnarray}
and
\begin{eqnarray}
(w\psi^{6}a^{3}n)_{,t}+\frac{1}{\sqrt{\eta}}
(\sqrt{\eta}w\psi^{6}a^{3}nv^{k})_{,k}=0.
\label{eq:SS99_2.8}
\end{eqnarray}

We can define the background Hubble parameter $H_{b}$ as
$
 H_{b}:={\dot{a}}/{a}$.
The scale factor $a(t)$ will be that of 
the flat FRW universe in the next section,
although the field equations up to here are independent from this choice.

\section{Long-wavelength scheme \label{sec:long_wavelength_limit}}

We here review the long-wavelength scheme based on 
Lyth {\it et al.}~\cite{Lyth:2004gb}.

\subsection{Basic assumptions}

To be precise, we focus on some fixed time
and put a fictitious parameter $\epsilon$ in front of the spatial partial 
derivative, e.g., $\partial_{i} \to \epsilon\partial_{i}$.
We expand exact solutions in a power series of $\epsilon$, require
the field equations at each order and 
finally set $\epsilon=1$. 
This scheme is called the gradient expansion.
We assume that the spacetime is approximately smooth 
for scales greater than $k^{-1}$ in terms of the comoving coordinate.
This implies that the physical scale of smoothing is $L=a/k$. Since the Hubble length $H_{b}^{-1}$ 
is the only natural 
scale of the cosmological evolution, we can make the identification 
$\epsilon=H_{b}^{-1}/L=k/(aH_{b})$. Thus, the gradient expansion 
implies that we assume that the smoothing length $L$ 
is much larger than the Hubble length, i.e., $L\gg
H_{b}^{-1}$. 

Based on the above consideration, we make two key assumptions.
First, we assume that $\psi$ is identically unity somewhere in the universe. 
This makes $a(t)$ be the scale factor for that part of the universe. 
Second, we assume that in the limit $\epsilon\to 0$, the universe
becomes locally homogeneous and isotropic, i.e., an FRW universe,
which we additionally assume is flat.
Thus, the measurable parts of the metric should reduce to the 
flat FRW one in the smoothing scale $L$ which is much larger than 
the Hubble length $H_{b}^{-1}$. 
This means that there exists a coordinate system with which 
the metric of any local region is written in the following form: 
\begin{equation}
 ds^{2}=-dt^{2}+a^{2}\eta_{ij}dx^{i}dx^{j}.
\label{eq:LMS07_8}
\end{equation}
Hence we assume $\beta^{i}=O(\epsilon)$, although this is a matter 
of coordinate choice.
As for the spatial metric, a homogeneous time-independent
$\tilde{\gamma}_{ij}$ can be transformed away, while a homogeneous
time-dependent $\tilde{\gamma}_{ij}$ should not exist by the 
present assumption. 
Since the term of $O(\epsilon)$ in $\dot{\tilde{\gamma}}_{ij}$
turns out to be decaying, we assume $\dot{\tilde{\gamma}}_{ij}=O(\epsilon^{2})$.
The above assumptions are partially justified in the literature in the 
context of inflationary cosmology~\cite{Jensen:1986nf,Sasaki:1998ug}.

\subsection{Energy conservation and curvature perturbation}

Let us consider a perfect fluid given by Eq.~(\ref{eq:perfect_fluid})
and choose the spatial coordinates so that the worldlines 
on which $x^{i}=\mbox{const}$ coincide with the worldlines of 
the fluid elements, which is called the comoving threading. This implies 
$ v^{i}=u^{i}/u^{t}=0$.
Then, $u^{\mu}$ is given by 
\begin{eqnarray}
 u^{\mu}&=&\left(\frac{1}{\sqrt{\alpha^{2}-\beta_{k}\beta^{k}}},0,0,0\right)
=\left(\frac{1}{\alpha},0,0,0\right)+O(\epsilon^{2}), \\
 u_{\mu}&=&\left(-\sqrt{\alpha^{2}-\beta^{k}\beta_{k}},\frac{\beta_{i}}{\sqrt{\alpha^{2}-\beta^{k}\beta_{k}}}\right)=\left(-\alpha,\frac{\beta_{i}}{\alpha}\right)+O(\epsilon^{2}).
\end{eqnarray}
The expansion of $u^{\mu}$ is given by
\begin{equation}
\theta:= u^{\mu}_{;\mu}=\frac{1}{\alpha
 \psi^{6}a^{3}}\partial_{t}
\left(\frac{\alpha
 \psi^{6}a^{3}}{\sqrt{\alpha^{2}-\beta^{i}\beta_{i}}}\right)
=\frac{3}{\alpha}\frac{(\psi^{2}a)_{,t}}{\psi^{2}a}+O(\epsilon^{2}),
\label{eq:LMS07_15_19}
\end{equation}
where it should be noted that $\tilde{\gamma}$ is time independent.
The relation between $t$ and the proper time $\tau$ along the worldlines
of the fluid elements is given by
\begin{equation}
 \frac{dt}{d\tau}=u^{t}=\frac{1}{\sqrt{\alpha^{2}-\beta^{i}\beta_{i}}}.
\end{equation}
The energy conservation law 
\begin{equation}
 0=-u_{\mu}T^{\mu\nu}_{;\nu}=\frac{d\rho}{d\tau}+(\rho+p)\theta
\label{eq:LMS07_17}
\end{equation}
implies 
\begin{equation}
 \frac{\dot{a}}{a}+2\frac{\dot{\psi}}{\psi}=-\frac{1}{3}\frac{\dot{\rho}}{\rho+p}+O(\epsilon^{2}).
\end{equation}
If we choose the {\it uniform-density slicing}, on which
$\rho=\mbox{const}$, 
and assume that the pressure 
is homogeneous to $O(\epsilon)$,  
$\dot{\psi}/\psi$ is also homogeneous to
$O(\epsilon)$. Since we assume that $\psi$ is identically unity at some point, 
we can conclude that $\psi$ is time independent to $O(\epsilon)$, i.e., 
\begin{equation}
 \dot{\psi}=O(\epsilon^{2}).
\label{eq:LMS07_24}
\end{equation} 

The expansion of $n^{\mu}$ is given by 
\begin{equation}
\theta_{n}:= n^{\mu}_{;\mu}=\frac{3}{\alpha}
\frac{(\psi^{2}a)_{,t}}{\psi^{2}a}-\frac{1}{\alpha
\psi^{6}a^{3}\sqrt{\eta}}\partial_{i}
\left(\sqrt{\eta}\psi^{6}a^{3}\beta^{i}\right)
\label{eq:LMS07_20}
\end{equation}
and this is related to the trace of the extrinsic curvature through 
\begin{equation}
 \theta_{n}=-K.
\label{eq:theta_n=-K}
\end{equation}
Since $\beta^{i}=O(\epsilon)$, Eqs.~(\ref{eq:LMS07_15_19}) and
(\ref{eq:LMS07_20}) imply
\begin{equation}
\theta=\theta_{n}+O(\epsilon^{2}).
\label{eq:theta=theta_n} 
\end{equation} 
This shows the equivalence between $\theta$ and $\theta_{n}$ to $O(\epsilon)$.
For convenience, we introduce the Hubble parameter $\tilde{H}$ as 
\begin{equation}
 \tilde{H}:= \frac{1}{3}\theta_{n}=\frac{1}{\alpha}\frac{(\psi^{2}a)_{,t}}{\psi^{2}a}+O(\epsilon^{2}).
\label{eq:LMS07_21}
\end{equation}
From Eqs.~(\ref{eq:LMS07_15_19}), (\ref{eq:LMS07_17}), (\ref{eq:theta=theta_n}) and
(\ref{eq:LMS07_21}), we find  
\begin{equation}
 \frac{d\rho}{d\tau}=-3\tilde{H}(\rho+p)+O(\epsilon^{2}).
\label{eq:LMS07_22}
\end{equation}

\subsection{Equivalence among different time slicings \label{subsec:equivalence_slicings}}

We use the following terminology about the slicing conditions.
We call the slicing which is orthogonal to the fluid worldlines,
namely, $n^{\mu}=u^{\mu}$,
the {\it comoving slicing}. This is independent of the comoving threading 
with which the worldline of the fluid coincides with that of the 
constant spatial coordinates.
We call the slicing on which 
the trace of the extrinsic curvature is uniform the 
{\it constant mean curvature (CMC) slicing}. Although this is sometimes 
called the uniform-Hubble slicing, we here avoid this terminology because of 
the ambiguity in the definition of the Hubble parameter in inhomogeneous 
cosmology. 
We call the slicing with $\alpha=1$ the {\it geodesic slicing}.
 
The comoving slicing implies $u_{i}=0$ and, hence, $J_{i}=0$.
This is possible only if $u^{\mu}$ is vorticity free because of the 
Frobenius theorem. This condition is physically 
reasonable in the early universe 
because the vorticity is not 
generated from vacuum fluctuation during inflation and 
because the vorticity is conserved for a perfect fluid in any spacetime.
Therefore,  in the comoving slicing, 
the momentum constraint (\ref{eq:SS99_2.21}) 
implies
\begin{equation}
 \partial_{i}\tilde{H}=O(\epsilon^{3})
\end{equation}
and the Hamiltonian constraint
(\ref{eq:SS99_2.20}) gives
\begin{equation}
 \tilde{H}^{2}=\frac{8\pi}{3}\rho+O(\epsilon^{2}),
\end{equation}
implying 
\begin{equation}
 \partial_{i}\rho=O(\epsilon^{3}).
\end{equation}
That is $\delta\rho=O(\epsilon^{2})$ and
$\delta\tilde{H}=O(\epsilon^{2})$ in the comoving slicing, 
where $\delta\rho$ and 
$\delta\tilde{H}$ are the inhomogeneous parts of
$\rho$ and $\tilde{H}$, respectively.
This means that the uniform-density, CMC and comoving slicings
coincide to $O(\epsilon)$.
This also guarantees that $\psi$ is time independent 
to $O(\epsilon)$ in each of those slicings and 
the time-dependent part appears only from $O(\epsilon^{2})$. 

From Eq.~(\ref{eq:LMS07_22}), we find 
\begin{equation}
 \frac{1}{\alpha}\dot{\rho}=-3\tilde{H}(\rho+p)+O(\epsilon^{2}).
\end{equation}
Since $\rho$ is homogeneous to $O(\epsilon)$, $\dot{\rho}$ 
is also homogeneous to $O(\epsilon)$. Hence, 
the lapse function is given by 
\begin{equation}
 \alpha=\frac{A(t)}{\rho+p}+O(\epsilon^{2}).
\end{equation}
where $A(t)$ is a function of $t$.
If $p$ is homogeneous to $O(\epsilon)$, 
which is the case for the barotropic 
equation of state, $\alpha$ is also homogeneous to $O(\epsilon)$
and we can choose $\alpha=1$ by rescaling the time 
coordinate.
Thus, the geodesic slicing is also equivalent to the uniform-density
slicing to $O(\epsilon)$.

\section{Long-wavelength solutions \label{sec:long_wavelength_solutions}}

\subsection{Gradient expansion}

Here we derive the expansion of the physical quantities based on the 
assumption of $\beta^{i}=O(\epsilon)$ and
$\dot{\tilde{\gamma}}_{ij}=O(\epsilon^{2})$.
In the limit $\epsilon\to 0$, the metric functions 
$\alpha$ and $\psi$ are assumed to be locally homogeneous. 
We can use the scaling of time so that $\alpha=1$ locally
in this limit and this suits the metric form 
given by Eq.~(\ref{eq:LMS07_8}). 

From Eq.~(\ref{eq:SS99_2.11}), we find
$\tilde{A}_{ij}=O(\epsilon^{2})$. 
We are still allowed to take general time 
slicings on which the density is uniform to $O(\epsilon)$.
As inhomogeneity appears in the mean curvature $K=-\theta_{n}$ from
$O(\epsilon^{2})$, $J_{i}=O(\epsilon^{3})$ follows 
from Eq.~(\ref{eq:SS99_2.21_variant}).
Combining this with Eq.~(\ref{eq:J_i}), 
we find $v_{i}+\beta_{i}=O(\epsilon^{3})$.
Therefore, 
since $\beta^{i}=O(\epsilon)$, we find $v^{i}=O(\epsilon)$.
Since $\dot{\tilde{\gamma}}_{ij}=O(\epsilon^{2})$, we can 
assign $h_{ij}=O(\epsilon^{2})$ using an appropriate coordinate
transformation, where $h_{ij}:=\tilde{\gamma}_{ij}-\eta_{ij}$.

In summary, if we assume 
\begin{eqnarray}
\beta^{i}=O(\epsilon), \quad 
\dot{\tilde{\gamma}}_{ij}=O(\epsilon^{2}),
\end{eqnarray}
then we deduce
\begin{eqnarray}
\psi=\psi(x^{k})=O(\epsilon^{0}), ~
v^{i}=O(\epsilon),~
\delta =O(\epsilon^{2}),~
  \tilde{A}_{ij}=O(\epsilon^{2}), ~h_{ij}=O(\epsilon^{2}),~
  \chi=O(\epsilon^{2}), ~\kappa =O(\epsilon^{2}),
\end{eqnarray}
where 
\begin{eqnarray}
  \delta:= \frac{\rho-\rho_{b}}{\rho_{b}},~
  h_{ij}:= \tilde{\gamma}_{ij}-\eta_{ij},~ 
  \chi:= \alpha-1, ~
  \kappa:=  \frac{K-K_{b}}{K_{b}}.
\end{eqnarray}
For later convenience, we define 
\begin{eqnarray}
 \delta_{n}:=\frac{n-n_{b}}{n_{b}}=\frac{1}{1+R}\delta+O(\epsilon^{4}),~ 
 \delta_{p}:=\frac{p-p_{b}}{p_{b}}=\frac{c_{s}^{2}}{R}\delta+O(\epsilon^{4}),
\label{eq:delta_n_delta_p} 
\end{eqnarray}
where we have used Eq.~(\ref{eq:Energy_conservation}) and defined
\begin{eqnarray}
R:=\frac{p_{b}}{\rho_{b}}, ~
c_{s}^{2}:=\left(\frac{dp}{d\rho}\right)_{b}. 
\label{eq:R_cs2}
\end{eqnarray}

Equations (\ref{eq:Energy_eq}), (\ref{eq:laplacianpsi}),
(\ref{eq:SS99_2.13}), (\ref{eq:SS99_2.14}) and (\ref{eq:SS99_2.8}) 
in $O(\epsilon^{0})$ give 
\begin{eqnarray}
 &&K_{b}=-3\frac{\dot{a}}{a}, \\
 &&\left(\frac{\dot{a}}{a}\right)^{2}=\frac{8\pi}{3}\rho_{b}, 
\label{eq:Friedmann_eq}\\
 &&\frac{\ddot{a}}{a}=-\frac{4\pi}{3}(\rho_{b}+3p_{b}), \\
 &&\dot{\rho}_{b}=-3(\rho_{b}+p_{b})\frac{\dot{a}}{a}, \\
 &&(a^{3}n_{b})_{,t}=0.
\label{eq:background_conservation}
\end{eqnarray}
This is a complete set of the Friedmann equations with spatially flat geometry.

To discuss the next order term in $\psi$, we decompose it into 
the following form:
\begin{equation}
 \psi(t,x^{k})=\Psi(x^{k})(1+\xi  (t,x^{k})),
\label{eq:psi_decomposition}
\end{equation}
where $\Psi=O(\epsilon^{0})$ and $\xi =O(\epsilon^{2})$.
It is also useful to note 
$
 w
=1+O(\epsilon^{6})
$ from Eq.~(\ref{eq:w}).

\subsection{Field equations in general gauge}
As we have seen in the previous section, 
the comoving, CMC, uniform-density and 
geodesic slicings coincide 
to $O(\epsilon)$. This ensures that we can consistently apply the 
gradient expansion in any of these four slicings. 
Here we derive the field equations in general gauge which is consistent with 
the long-wavelength scheme by generalizing the formulation by
Shibata and Sasaki~\cite{Shibata:1999zs}.

First, Eq.~(\ref{eq:SS99_2.13}) results in
$\dot{\Psi}=0$.
This is the starting point of the formulation.
Then, Eqs.~(\ref{eq:Energy_eq_decomposition}) and 
(\ref{eq:SS99_2.8}) yield
\begin{equation}
 \dot{\delta}+6\dot{\xi }+\nabla_{k}v^{k}-3H_{b}R(\delta-\delta_{p}-\chi-\kappa)=O(\epsilon^{4}),
\label{eq:Energy_eq_perturbation}
\end{equation}
and
\begin{equation}
 \dot{\delta}_{n}+6\dot{\xi }+\nabla_{k}v^{k}=O(\epsilon^{4}),
\label{eq:SS99_3.2}
\end{equation}
respectively, where 
\begin{eqnarray}
 \nabla_{k}v^{k}:=\frac{1}{\Psi^{6}\eta^{1/2}}\partial_{k}(\eta^{1/2}\Psi^{6}v^{k}).
\end{eqnarray}
Equations~(\ref{eq:SS99_2.9}),  
(\ref{eq:SS99_2.13}), (\ref{eq:laplacianpsi}),
(\ref{eq:SS99_2.14}), (\ref{eq:SS99_2.11}), (\ref{eq:SS99_2.12}), 
(\ref{eq:SS99_2.21_variant}) 
yield
\begin{eqnarray}
&&\partial_{t}[a^{3}(1+R)\rho_{b}u_{j}]=-a^{3}\rho_{b}
\left[R\partial_{j}\delta_{p}+(1+R)\partial_{j}\chi\right]+O(\epsilon^{5}).
\label{eq:SS99_3.3_variant}, \\
&& 6\dot{\xi }-3H_{b}(\chi+\kappa)-\nabla_{k}\beta^{k}
=O(\epsilon^{4}), 
\label{eq:SS99_3.4} \\
&&
\bar{\Delta}\Psi=-2\pi \Psi^{5}a^{2}\rho_{b}(\delta-2\kappa)+O(\epsilon^{4}),
\label{eq:SS99_3.5}
\\
&& H_{b}^{-1}\dot{\kappa} = 
\frac{1}{2}(3R-1)\kappa-\frac{3}{2}(1+R)\chi-\frac{1}{2}(\delta+3R\delta_{p})+O(\epsilon^{4}),
\label{eq:SS99_3.12} \\
&& \partial_{t}h_{ij}=-2\tilde{A}_{ij}+\eta_{ik}\bar{\cal D}_{j}\beta^{k}
+\eta_{jk}\bar{\cal D}_{i}\beta^{k}-\frac{2}{3}\eta_{ij}\bar{\cal
D}_{k}\beta^{k}+O(\epsilon^{4}), 
\label{eq:SS99_3.7}\\
&& \partial_{t}\tilde{A}_{ij}+\frac{3\dot{a}}{a}\tilde{A}_{ij}=\frac{1}{a^{2}\Psi^{4}}
\left[-\frac{2}{\Psi}\left(\bar{\cal D}_{i}\bar{\cal D}_{j}\Psi-\frac{1}{3}\eta_{ij}\bar{\Delta}\Psi\right)+\frac{6}{\Psi^{2}}\left(\bar{\cal D}_{i}\Psi\bar{\cal D}_{j}\Psi-\frac{1}{3}\eta_{ij}\bar{\cal D}^{k}\Psi\bar{\cal D}_{k}\Psi\right)\right]+O(\epsilon^{4}),
\label{eq:SS99_3.8}\\
&& \bar{\cal D}_{i}(\Psi^{6}\tilde{A}^{i}_{j})+2H_{b}\Psi^{6}\bar{\cal D}_{j}\kappa=8\pi\Psi^{6}(1+R)\rho_{b}u_{j}
+O(\epsilon^{5}), 
\label{eq:momentum_constraint_CMC_approx}
\end{eqnarray}
respectively, where we have used
\begin{eqnarray}
&&E+S^{k}_{k}=\rho+3p+O(\epsilon^{6}), \\
&&S_{ij}-\frac{\gamma_{ij}}{3}S^{k}_{k}=O(\epsilon^{6})
\end{eqnarray}
following from Eqs.~(\ref{eq:E})--(\ref{eq:S_ij}).
Equations (\ref{eq:Energy_eq_perturbation}), (\ref{eq:SS99_3.2}) 
and (\ref{eq:SS99_3.4}) give
\begin{eqnarray}
&& \dot{\delta}-\dot{\delta}_{n}-3H_{b}R(\delta-\delta_{p}-\chi-\kappa)=O(\epsilon^{4}),
\label{eq:dotdelta-dotdeltan} \\
&&
 \dot{\delta}-3H_{b}R(\delta-\delta_{p})+3(1+R)H_{b}(\chi+\kappa)=O(\epsilon^{4}), 
\label{eq:SS99_3.11_variant}\\
 &&\dot{\delta}-(1+R)\dot{\delta}_{n}-3H_{b}R(\delta-\delta_{p})=O(\epsilon^{4}),
\end{eqnarray}
where it should be noted that
$
\nabla_{k}(\beta^{k}+v^{k})=O(\epsilon^{4})
$.

\subsection{Long-wavelength solutions \label{subsec:long_wavelength_solutions}}

\subsubsection{$\tilde{A}_{ij}$ and $h_{ij}$}
We can see that $\tilde{A}_{ij}$ and $h_{ij}$
are not sensitive to the choice of time slicings
to $O(\epsilon^{2})$.
Equation (\ref{eq:SS99_3.8}) admits the following solution for 
$\tilde{A}_{ij}$:
\begin{equation}
 \tilde{A}_{ij}=p_{ij}
\frac{1}{a^{3}}\left[\int_{0}^{a}\frac{d\tilde{a}}{H_{b}(\tilde{a})}\right]+O(\epsilon^{4}),
\end{equation}
where we have put
\begin{equation}
 p_{ij}(x^{k}):=\frac{1}{\Psi^{4}}
\left[
-\frac{2}{\Psi}\left(\bar{\cal D}_{i}\bar{\cal D}_{j}\Psi-\frac{1}{3}\eta_{ij}\bar{\Delta}\Psi\right)+\frac{6}{\Psi^{2}}\left(\bar{\cal D}_{i}\Psi\bar{\cal D}_{j}\Psi-\frac{1}{3}\eta_{ij}\bar{\cal D}^{k}\Psi\bar{\cal D}_{k}\Psi\right)
\right]
\end{equation}
and we have omitted the integration constant because it generally 
gives a decaying mode.

The expression of $h_{ij}$ depends on the choice of the spatial
coordinates. For the normal coordinates, where $\beta^{i}=0$, 
by integrating Eq.~(\ref{eq:SS99_3.7}), we find 
\begin{eqnarray}
 h_{ij}=-2p_{ij}\int_{0}^{a}\frac{d\tilde{a}}{\tilde{a}^{4}H_{b}(\tilde{a})}\int_{0}^{\tilde{a}}
\frac{d\bar{a}}{H_{b}(\bar{a})}+C_{ij}+O(\epsilon^{4}).
\end{eqnarray}
We drop the constant tensor $C_{ij}$ so that $h_{ij}\to 0$ 
in the limit $t\to 0$ by choosing
appropriate spatial coordinates.
Of course, other coordinate conditions may be useful.
For example, in spherical symmetry, we can even have $h_{ij}=0$
identically, 
for which $\beta^{i}$ is accordingly determined. This 
is called the conformally flat spatial coordinates.

If we assume $\Gamma=\mbox{const}$, we have the background solution
from Eq.~(\ref{eq:Friedmann_eq}) as
\begin{equation}
 a=a_{0}\left(\frac{t}{t_{0}}\right)^{\frac{2}{3\Gamma}},
\label{eq:power_law_FRW}
\end{equation}
and we can explicitly write $\tilde{A}_{ij}$ and $h_{ij}$ 
in the following form:
\begin{eqnarray}
 \tilde{A}_{ij}&=&
\frac{2}{3\Gamma+2}p_{ij}H_{b}
\left(\frac{1}{aH_{b}}\right)^{2}+O(\epsilon^{4}), 
\label{eq:A_tilde_ij_solution}\\
 h_{ij}&=&-\frac{4}{(3\Gamma+2)(3\Gamma-2)}p_{ij}\left(\frac{1}{aH_{b}}\right)^{2}+O(\epsilon^{4}), 
\label{eq:h_ij_solution}
\end{eqnarray}
where the bottom one is obtained in the normal coordinates.
$\tilde{A}_{ij}$ and $h_{ij}$ depend on time as
\begin{eqnarray}
 \tilde{A}_{ij}\propto  t^{1-\frac{4}{3\Gamma}}
\quad
\mbox{and}
\quad  
 h_{ij}\propto t^{2-\frac{4}{3\Gamma}}, 
\end{eqnarray}
respectively.

\subsubsection{CMC slicing \label{subsec:CMC_S}}
Hereafter we assume $\Gamma=\mbox{const}$ for simplicity. 
Then, Eqs.~(\ref{eq:delta_n_delta_p}) and (\ref{eq:R_cs2}),  
respectively, yield
\begin{eqnarray}
 \delta_{n}&=&\frac{1}{\Gamma}\delta+O(\epsilon^{4}), \\
 \delta_{p}&=&\delta+O(\epsilon^{4}).
\end{eqnarray}
We present the solutions in the CMC slicing, where $\kappa=0$.
Equation~(\ref{eq:SS99_3.5}) is solved for $\delta$ to give
\begin{equation}
  \delta=-\frac{\bar{\Delta}\Psi}{2\pi \Psi^{5}a^{2}\rho_{b}}+O(\epsilon^{4}).
\end{equation}
Defining
\begin{equation}
 f(x^{k}):=-\frac{4}{3}\frac{\bar{\Delta}\Psi}{\Psi^{5}}
\label{eq:definition_f}
\end{equation}
and using Eqs.~(\ref{eq:Friedmann_eq}) and (\ref{eq:SS99_3.12}), we can express $\delta$ and $\chi$ as follows:
\begin{eqnarray}
 \delta&=&f\left(\frac{1}{a H_{b}}\right)^{2}+O(\epsilon^{4}), 
\label{eq:delta_f_CMC}\\
 \chi&=&-\frac{3\Gamma-2}{3\Gamma}f\left(\frac{1}{a H_{b}}\right)^{2}+O(\epsilon^{4}).
\end{eqnarray}
Then, from Eq.~(\ref{eq:SS99_3.3_variant}), we find 
\begin{equation}
 u_{j}=\frac{1}{3\Gamma}\partial_{j}f\frac{1}{a}\left[\int^{a}_{0}\frac{d\tilde{a}}{H_{b}(\tilde{a})}+C\right]\left(\frac{1}{a H_{b}}\right)^{2}+O(\epsilon^{5}),
\end{equation}
where $C$ is a constant of integration. If $C$ is nonvanishing, 
there appears a decaying mode for $1\le \Gamma<2$. 
So we assume $C=0$.
Thus, we have
\begin{equation}
 u_{j}=u^{t}(v_{j}+\beta_{j})=\frac{1}{3\Gamma}\partial_{j}f \frac{1}{a}\left[\int^{a}_{0}\frac{d\tilde{a}}{H_{b}(\tilde{a})}\right]\left(\frac{1}{a H_{b}}\right)^{2}
+O(\epsilon^{5}).
\label{eq:u_j_lwll}
\end{equation}
We can explicitly prove that the above 
solves the momentum constraint equation 
(\ref{eq:momentum_constraint_CMC_approx}) 
to $O(\epsilon^{3})$ using the identity
\begin{equation}
 \bar{\cal D}^{j}(\Psi^{6}p_{ij} )=\Psi^{6}\bar{\cal D}_{i}f .
\end{equation}

If we choose the normal coordinates, where $\beta^{i}=0$, 
$v_{j}$ is obtained by Eq.~(\ref{eq:u_j_lwll}) as
\begin{equation}
 v_{j}=\frac{1}{3\Gamma}\partial_{j}f \frac{1}{a}\left[\int^{a}_{0}\frac{d\tilde{a}}{H_{b}(\tilde{a})}\right]\left(\frac{1}{a H_{b}}\right)^{2}+O(\epsilon^{5}).
 \end{equation}
In this case, $\xi $ is obtained by integrating Eq.~(\ref{eq:SS99_3.4}) as
\begin{equation}
 \xi = -\frac{1}{6\Gamma}f (x^{k})\left(\frac{1}{aH_{b}}\right)^{2}
+O(\epsilon^{4}), 
\end{equation}
where the integration constant is absorbed into $\Psi$.

Using the background solution (\ref{eq:power_law_FRW}),
we find
\begin{eqnarray}
 \delta&=&f \left(\frac{1}{a H_{b}}\right)^{2}+O(\epsilon^{4}), 
\label{eq:delta_CMC}\\
 \kappa&=&0 ,\\
 \chi&=&-\frac{3\Gamma-2}{3\Gamma}f \left(\frac{1}{a H_{b}}\right)^{2}+O(\epsilon^{4}), \\
 u_{j}&=&u^{t}(v_{j}+\beta_{j})=\frac{2}{3\Gamma(3\Gamma+2)}\partial_{j}f 
a\left(\frac{1}{a H_{b}}\right)^{3}+O(\epsilon^{5}), \label{eq:ur_CMC}\\
 v_{j}&=& \frac{2}{3\Gamma(3\Gamma+2)}\partial_{j}f  a
  \left(\frac{1}{aH_{b}}\right)^{3}+O(\epsilon^{5}), 
\label{eq:v_j_CMC} \\
 \xi &=& -\frac{1}{6\Gamma}f
  (x^{k})\left(\frac{1}{aH_{b}}\right)^{2}+O(\epsilon^{4}),
\label{eq:xi_CMC}
\end{eqnarray}
where the last two equations are obtained in the normal coordinates.
The time dependence of the perturbations is summarized as follows:
\begin{eqnarray}
 \delta\propto t^{2-\frac{4}{3\Gamma}},~
 \chi\propto t^{2-\frac{4}{3\Gamma}}, ~
 u_{j}=u^{t}(v_{j}+\beta_{j})\propto  t^{3-\frac{4}{3\Gamma}}, ~
 \xi \propto   t^{2-\frac{4}{3\Gamma}} , ~
 v_{j}\propto  t^{3-\frac{4}{3\Gamma}},
\end{eqnarray}
where the last two equations are obtained 
in the normal coordinates.

\subsubsection{Uniform-density slicing \label{subsec:UD_S}}
The uniform density slicing implies $\delta=0$. By the assumption $\Gamma=$const, we find
$\delta_{n}=\delta_{p}=0$.  
From Eqs.~(\ref{eq:SS99_3.5}), (\ref{eq:SS99_3.11_variant}) and (\ref{eq:SS99_3.3_variant}), we find 
\begin{eqnarray}
 \kappa&=&-\frac{1}{2}f \left(\frac{1}{a H_{b}}\right)^{2}+O(\epsilon^{4}), \\
 \chi&=&\frac{1}{2}f \left(\frac{1}{a H_{b}}\right)^{2}+O(\epsilon^{4}), \\
 u_{j}&=&u^{t}(v_{j}+\beta_{j})=-\frac{1}{2}\partial_{j}f \frac{1}{a}\left[\int^{a}_{0}\frac{d\tilde{a}}{H_{b}(\tilde{a})}\right]\left(\frac{1}{a H_{b}}\right)^{2}
+O(\epsilon^{5}),
\end{eqnarray}
where $f $ is defined by Eq.~(\ref{eq:definition_f}).
We can explicitly show that the momentum constraint
(\ref{eq:momentum_constraint_CMC_approx}) is satisfied to this order.
Using the background solution (\ref{eq:power_law_FRW}), we have
\begin{eqnarray}
 \delta&=&0, \\
 \kappa&=&-\frac{1}{2}f \left(\frac{1}{a H_{b}}\right)^{2}+O(\epsilon^{4}), \\
 \chi&=&\frac{1}{2}f \left(\frac{1}{a H_{b}}\right)^{2}+O(\epsilon^{4}), \\
 v_{j}&=&-\frac{1}{3\Gamma+2}\partial_{j}f a\left(\frac{1}{a H_{b}}\right)^{3}+O(\epsilon^{5}), \\
 \xi &=&O(\epsilon^{4}),
\end{eqnarray}
where the last two expressions are obtained in the normal coordinates,
and in the bottom equation we have used that
$\xi $ can be shown to be time independent and is absorbed 
into $\Psi$ to $O(\epsilon^{2})$.
Therefore, in the uniform-density slicing with the normal coordinates, 
$\psi$ is time independent to $O(\epsilon^{2})$.

\subsubsection{Comoving slicing \label{subsec:C_S}}
In the comoving slicing, $u^{\mu}=n^{\mu}$ and hence
\begin{equation}
 u_{i}=0.
\end{equation}
Then, Eq.~(\ref{eq:SS99_3.3_variant}) implies
\begin{equation}
 (\Gamma-1)\delta+\Gamma\chi=O(\epsilon^{4}).
\end{equation}
Using this and Eqs.~(\ref{eq:SS99_3.12})
and (\ref{eq:dotdelta-dotdeltan}), we find 
\begin{eqnarray}
 \partial_{s}\kappa&=&\frac{1}{2}(3\Gamma-4)\kappa-\frac{1}{2}\delta+O(\epsilon^{4}), \\
 \partial_{s}\delta&=&-3\Gamma \kappa+3(\Gamma-1)\delta+O(\epsilon^{4}),
\end{eqnarray}
where we put $s=\ln a$. 
The matrix
\begin{eqnarray}
 \left(\begin{array}{cc}
  (3\Gamma-4)/2&-1/2 \\
  -3\Gamma & 3(\Gamma-1)
       \end{array}\right)
\end{eqnarray}
has eigenvalues 
$(3\Gamma-2)$
and 
$3(\Gamma-2)/2$
with associated eigenvectors
\begin{eqnarray}
 \left(\begin{array}{c}
  1\\
-3\Gamma\\
       \end{array}\right)
\quad \mbox{and} \quad 
 \left(\begin{array}{c}
  1\\
2\\
       \end{array}\right),
\end{eqnarray}
respectively.
Therefore, the general solution for $\kappa$ and $\delta$ is given by 
\begin{eqnarray}
 \left(
\begin{array}{c}
\kappa\\
\delta
       \end{array}
\right) 
=
C_{1} \left(
\begin{array}{c}
1\\
-3\Gamma
       \end{array}
\right)a^{3\Gamma-2} 
+C_{2} \left(
\begin{array}{c}
1\\
2
       \end{array}
\right)a^{\frac{3}{2}(\Gamma-2)}+O(\epsilon^{4}),
\label{eq:two_modes}
\end{eqnarray}
where $C_{1}$ and $C_{2}$ are integration constants.
For $2/3<\Gamma<2$, the first term grows, while 
the second term decays in time. Hence we drop the second term 
by choosing $C_{2}=0$.
Thus, we have
\begin{equation}
 \kappa=-\frac{1}{3\Gamma}\delta+O(\epsilon^{4}).
\end{equation}
Substituting this into Eq.~(\ref{eq:SS99_3.5}), we have
\begin{equation}
 \delta=\frac{3\Gamma}{3\Gamma+2}f 
\left(\frac{1}{aH_{b}}\right)^{2},
\label{eq:delta_CC}
\end{equation}
where $f$ is defined by Eq.~(\ref{eq:definition_f}).
It should be noted that the time dependence of $\delta$ is consistent
with the first term on the right-hand side of Eq.~(\ref{eq:two_modes}).
Now it is straightforward to see that $\kappa$, $\chi$ and $\xi $ are given
by 
\begin{eqnarray}
 \kappa&=&-\frac{1}{3\Gamma+2}f \left(\frac{1}{aH_{b}}\right)^{2}+O(\epsilon^{4}), \\
 \chi&=&-\frac{3(\Gamma-1)}{3\Gamma+2}f
  \left(\frac{1}{aH_{b}}\right)^{2}+O(\epsilon^{4}), 
\label{eq:alpha2_CC}\\
 u_{j}&=& 0, \\
 v_{j}&=& 0, \\
 \xi &=&-\frac{1}{2(3\Gamma+2)}f
  \left(\frac{1}{aH_{b}}\right)^{2}+O(\epsilon^{4}),
\label{eq:xi_CC} 
\end{eqnarray}
where the last two expressions are obtained 
in the normal coordinates.
We can explicitly show that the momentum constraint
(\ref{eq:momentum_constraint_CMC_approx}) is satisfied to this order.
The combination of the comoving slicing and comoving threading, 
i.e., $u_i=\beta_{i}=0$, is called the comoving gauge.

\subsubsection{Geodesic slicing \label{subsec:G_S}}
In the geodesic slicing, it is straightforward to obtain
\begin{eqnarray}
 \delta&=&\frac{3\Gamma}{9\Gamma-4}f \left(\frac{1}{aH_{b}}\right)^{2}+O(\epsilon^{4}), \\
 \kappa&=&-\frac{3\Gamma-2}{9\Gamma-4}f \left(\frac{1}{aH_{b}}\right)^{2}+O(\epsilon^{4}), \\
 \chi&=&0, \\
 u_{j}&=&-\frac{6(\Gamma-1)}{(9\Gamma-4)(3\Gamma+2)}\partial_{j}f a\left(\frac{1}{a H_{b}}\right)^{3}
+O(\epsilon^{5}), \\
 v_{j}&=&-\frac{6(\Gamma-1)}{(9\Gamma-4)(3\Gamma+2)}\partial_{j}f a\left(\frac{1}{a H_{b}}\right)^{3}
+O(\epsilon^{5}), \\
 \xi &=&-\frac{1}{2(9\Gamma-4)}f \left(\frac{1}{aH_{b}}\right)^{2}+O(\epsilon^{4}),  
\end{eqnarray} 
where the last two expressions are obtained in the normal coordinates.
The gauge condition of $\alpha=1$ and $\beta^{i}=0$,
i.e., the geodesic slicing with the normal coordinates, 
is called the synchronous gauge.

\subsubsection{Relations among the different time slicings}

In Sec.~\ref{sec:long_wavelength_limit}, we 
have seen that the slicings in the long-wavelength scheme coincide 
to $O(\epsilon)$. This means that the zeroth-order 
curvature variable $\Psi=\Psi(x^{k})$ should be common, i.e., 
\begin{equation}
 \Psi_{\rm CMC}=\Psi_{\rm UD}=\Psi_{C}=\Psi_{G}
\end{equation}
for the same spacetime, where CMC, UD, C and G stand for 
the CMC slicing, uniform-density slicing,  
comoving slicing, and geodesic slicing, respectively.
Using the results obtained in Sec.~\ref{subsec:long_wavelength_solutions},
we can find the relations among the perturbations in the different 
slicings.
Irrespective of the choice of $\beta^{i}$, we can conclude 
\begin{equation}
 \tilde{A}_{ij, {\rm CMC}}=\tilde{A}_{ij, {\rm UD}}+O(\epsilon^{4})=\tilde{A}_{ij,C}+O(\epsilon^{4})=\tilde{A}_{ij,G}+O(\epsilon^{4})
\end{equation}
and 
\begin{eqnarray}
 \delta_{C}&=&\frac{3\Gamma}{3\Gamma+2}\delta_{\rm CMC}+O(\epsilon^{4}), 
  \delta_{\rm UD}=0,   \delta_{G}=\frac{3\Gamma}{9\Gamma-4}\delta_{\rm CMC}, 
\label{eq:density_perturbation_four_slicings}\\
 \kappa_{\rm CMC}&=&0,  \kappa_{C}=\frac{2}{3\Gamma+2}\kappa_{\rm UD}+O(\epsilon^{4}),  
   \kappa_{G}=\frac{2(3\Gamma-2)}{9\Gamma-4}\kappa_{\rm UD}\\
 \chi_{C}&=&\frac{9\Gamma(\Gamma-1)}{(3\Gamma-2)(3\Gamma+2)}\chi_{\rm CMC}+O(\epsilon^{4}),  
 \chi_{\rm UD}=-\frac{3\Gamma}{2(3\Gamma-2)}\chi_{\rm CMC}+O(\epsilon^{4}),
   \chi_{G}=0, \\
 u_{j~C}&=&0, u_{j~{\rm UD}}=-\frac{3\Gamma}{2}u_{j~{\rm CMC}}+O(\epsilon^{4}), 
  u_{j~G}=-\frac{9\Gamma(\Gamma-1)}{9\Gamma-4}u_{j~{\rm CMC}}.
\end{eqnarray}
In the normal coordinates, where $\beta^{i}=0$, 
we can also conclude the following relations:
\begin{eqnarray}
 h_{ij, {\rm CMC}}&=&h_{ij,{\rm
  UD}}+O(\epsilon^{4})=h_{ij,C}+O(\epsilon^{4})=h_{ij,G}+O(\epsilon^{4}), \\
 v_{j~{\rm UD}}&=&-\frac{3\Gamma}{2}v_{j~{\rm CMC}}+O(\epsilon^{4}), 
  v_{j~C}=0, 
  v_{j~G}=-\frac{9\Gamma(\Gamma-1)}{9\Gamma-4}v_{j~{\rm CMC}}, \\
 \xi _{C}&=&\frac{3\Gamma}{3\Gamma+2}\xi _{\rm CMC}+O(\epsilon^{4}), 
  \xi _{\rm UD}=O(\epsilon^{4}),   \xi _{G}=\frac{3\Gamma}{9\Gamma-4}\xi _{\rm CMC}.
\end{eqnarray}

\section{Spherically symmetric spacetimes}

One of the important motivations of the following sections
is to compare the results of numerical 
simulations in two very different approaches to PBH formation
in spherical symmetry.
The one uses the CMC slicing with the conformally flat spatial coordinates,
while the other uses the comoving slicing with the comoving threading,
which is called the Misner-Sharp formulation of the Einstein 
equation in spherical symmetry with a perfect fluid.
For this reason, we review the two formulations of the Einstein
equations in spherical symmetry. 

\subsection{Gauge conditions in the two approaches}

Shibata and Sasaki~\cite{Shibata:1999zs} adopt the conformally flat spatial coordinates.  
The line element is given by~\cite{footnote1}
\begin{equation}
 ds^{2}=-(\alpha^{2}-\psi^{4}a^{2}\beta^{2}r^{2})dt^{2}+2\psi^{4}a^{2}\beta
  rdrdt
+\psi^{4}a^{2}(dr^{2}+r^{2}d\Omega^{2}),
\label{eq:metric_spherical_SS}
\end{equation}
where $d\Omega^{2}=d\theta^{2}+\sin^{2}\theta d\phi^{2}$ is the 
line element on the unit two-sphere. 
Thence, we have
$\beta^{r}=r\beta$ and $\tilde{\gamma}_{ij}=\eta_{ij}$.
The spatial metric is conformally flat and this is also a minimal
distortion gauge because $\dot{\tilde{\gamma}}_{ij}=0$. 
As for the slicing condition, 
Shibata and Sasaki~\cite{Shibata:1999zs} adopt the CMC slicing 
on which $K=-3H_{b}(t)$ and 
formulated the initial value problem of the Einstein equation. 
The combination of the conformally flat 
spatial coordinates and the CMC slicing 
fixes the gauge.
Since the full set of field equations can be derived from the 
equations in Sec.~\ref{subsec:cosmological_conformal_decomposition}
and are given in Shibata and Sasaki~\cite{Shibata:1999zs}, we do not repeat them here.

If we adopt the comoving slicing with the comoving threading, 
we have a compact set of field equations and this is called the 
Misner-Sharp formulation~\cite{Misner:1964je}. 
The line element is given by 
\begin{equation}
 ds^{2}=-\alpha^{2}dt^{2}+b^{2}dr^{2}+R^{2}d\Omega^{2}.
\label{eq:spherically_symmetric_diagonal_metric}
\end{equation}
The following equations derive from Eqs.~(\ref{eq:Hamiltonian_constraint}), (\ref{eq:momentum_constraint}),
(\ref{eq:extrinsic_curvature}), (\ref{eq:evolution_equation}),
(\ref{eq:Energy_eq}), (\ref{eq:Euler_eq}), (\ref{eq:E}) and
(\ref{eq:J_i})
with $\beta^{i}=v^{i}=0$: 
\begin{eqnarray}
 M'&=&4\pi \rho  R^{2}R', 
\label{eq:M'}\\
\dot{M}&=&-4\pi p R^{2}\dot{R}, \\
 \dot{R}'&=& \dot{R}\displaystyle\frac{\alpha'}{\alpha}+R'\frac{\dot{b}}{b},\\
 p'&=&-(\rho +p)\displaystyle\frac{\alpha'}{\alpha} \\
 M&=&\displaystyle\frac{R}{2}\left[1-\displaystyle\frac{(R')^{2}}{b^{2}}+\frac{(\dot{R})^{2}}{\alpha^{2}}\right],
\label{eq:definition_M}
\end{eqnarray}
where the prime denotes the partial derivative with respect to $r$ and
$M$ is called the Misner-Sharp mass~\cite{Misner:1964je}.
This formulation has been adopted for PBH studies 
by many authors including Polnarev and Musco~\cite{Polnarev:2006aa}.

\subsection{Areal radius, mass excess and compaction function
\label{subsec:areal_radius_etc}}

In spherical symmetry, the areal radius is defined by $R=\sqrt{A/(4\pi)}$,
where $A$ is the area of the 2-sphere with constant $t$ and $r$.
We define the Kodama vector $K^{A}$ by
$
 K^{A}=\epsilon^{AB}\partial_{B}R
$,
where $\epsilon_{AB}$ is the totally antisymmetric tensor 
in the two-dimensional manifold charted by $t$ and $r$ 
and the Latin uppercase indices run over 0 and 1, and $\epsilon_{AB}$
is given by
$
 \epsilon_{AB}=\sqrt{-G}\varepsilon_{AB}
$
with the Levi-Civita symbol $\varepsilon_{AB}$
and the determinant $G$ of the two-dimensional metric $G_{AB}$.
We raise the indices of $\epsilon_{AB}$ as 
$\epsilon_{A}^{~B}=\epsilon_{AC}G^{CB}$
and $\epsilon^{AB}=G^{AC}\epsilon_{C}^{~B}$.
We trivially extend $K^{A}$ to $K^{\mu}$ as a vector field 
in the four-dimensional manifold.
Since $S^{\mu}=T^{\mu}_{\nu}K^{\nu}$ is a conserved current,
\begin{equation}
 M:=-\int S^{\mu}d\Sigma_{\mu}=-\int S^{\mu}n_{\mu}d\Sigma
=\int S^{t}\alpha\sqrt{\gamma}dx^{3}
\label{eq:Kodama_mass}
\end{equation}
is a conserved mass, where in the last expression
$\Sigma$ is chosen to the constant $t$ hypersurface
and the integration is done within a ball on $\Sigma$.
$M$ is called the Kodama mass. This is equivalent to the Misner-Sharp
mass in the present setting.

Shibata and Sasaki~\cite{Shibata:1999zs} introduce the notion of 
a compaction function. Here we define it in more general settings.
To this end we first define an excess $\delta M$ in the Kodama mass by the 
difference between the Kodama masses of the two spheres of same $t$ and {\it
areal radius} $R$ (not $r$) in two different spherically symmetric 
spacetimes. This definition of $\delta M$ is covariant with respect 
to the choice of spatial coordinates but does depend on the slicing.
We use the flat FRW spacetime as a reference spacetime to define the 
mass excess.

In an analogy with asymptotic flatness, if 
initial data approach those of the flat FRW one so fast that 
the mass excess has a finite limit as we take $r\to \infty$, 
we call such data {\it asymptotically flat FRW}. 
For asymptotically flat FRW data, 
if the mass excess approaches zero as we take $r\to\infty$, 
we shall call such initial data {\it compensated};  
otherwise, we shall call it as {\it uncompensated}. 
It is not clear which model is more physically realistic, compensated or
uncompensated. Clearly, uncompensated models are more generic than the 
compensated ones. Physics should be as generic as possible within
the framework of asymptotically flat FRW data.
On the other hand, uncompensated models might be regarded 
as perturbations of infinitely long wavelength.

We define a compaction function ${\cal C}(t,r)$ 
by the ratio of the mass excess 
to the common areal radius of the two spheres, i.e., 
\begin{equation}
 {\cal C}(t,r):=\frac{\delta M(t,r)}{R(t,r)}.
\end{equation}
By definition, ${\cal C}(t,r)=0$ for the flat FRW spacetime. For
asymptotically flat FRW data, we find 
$
 {\cal C}\propto 1/R
$
as we fix $t$ and take the limit $r\to \infty$. 
In the following we express the compaction function in terms of the density perturbation
first in the Misner-Sharp formulation and later in more general slicings.

\subsubsection{Comoving slicing}
In the Misner-Sharp formulation, for which the line element is given in
the form of Eq.~(\ref{eq:spherically_symmetric_diagonal_metric}), 
the components $K^{\mu}$ of the 
Kodama vector are given by 
\begin{eqnarray}
 K^{t}=-\frac{R'}{\alpha b}, \quad 
 K^{r}=\frac{\dot{R}}{\alpha b},\quad K^{\theta}=K^{\phi}=0.
\end{eqnarray}
Then, we find 
\begin{equation}
S^{t}=T^{t}_{\mu}K^{\mu}=T^{t}_{t}K^{t}=\frac{R'}{\alpha b}\rho,
\end{equation}
where we have used $u_{i}=0$ in this coordinate system, 
and hence
\begin{equation}
 M(t,r)=4\pi \int_{0}^{r} dx \rho R^{2}(t,x)R'(t,x).
\end{equation}
Note that this expression is fully valid even in the general 
spherically symmetric spacetime and shows the equivalence 
between the Kodama mass and 
the Misner-Sharp mass.
We denote the Kodama mass in the flat FRW spacetime by $M_F$, which is given by
\begin{equation}
 M_{F}(t,r)=4\pi \rho_{b} a^{3}\int_{0}^{r}dxx^{2}.
\end{equation} 
Then the mass excess is defined by 
\begin{equation}
 \delta M(t,r):=M(t,r)-M_{F}(t,\tilde{r}),
\end{equation}
where $\tilde{r}$ in the FRW spacetime gives the same areal radius $R(t,r)$ of the 
sphere of $r$ on the constant $t$ hypersurface in the perturbed spacetime, namely, 
\begin{equation}
a(t)\tilde{r}=R(t,r).
\end{equation}
Since $\Sigma$ is the constant $t$
hypersurface in both unperturbed and perturbed spacetimes, 
we find 
\begin{equation}
 a(t)d\tilde{r}=R'(t,r)dr,
\end{equation}
therefore, 
\begin{equation}
 M_{F}(t,\tilde{r})=4\pi \rho_{b}\int_{0}^{r}dx R^{2}(t,x)R'(t,x).
\end{equation} 
Thus, we find the following exact relation: 
\begin{equation}
 \delta M(t,r)=4\pi \rho_{b}\int^{r}_{0} dx  R^{2}(t,x)R'(t,x)\delta(t,x),
\end{equation}
where 
\begin{equation}
\delta:=\frac{\rho-\rho_{b}}{\rho_{b}}. 
\end{equation}
This is independent of the choice of the spatial coordinates.
Thus, we have established the theorem that in spherical symmetry
and in the comoving slicing, 
the density perturbation integrated over the perturbed spatial 
geometry exactly coincides with the mass excess.
We can also have the exact expression for the compaction function
as follows:
\begin{equation}
{\cal C}=\frac{1}{2}\bar{\delta}(H_{b}R)^{2},
\label{eq:C_bar_delta_comoving}
\end{equation} 
where the averaged density perturbation $\bar{\delta}$ is defined as
\begin{equation}
 \bar{\delta}(t,r):=\frac{\int \delta d\Sigma}{\int d\Sigma}=\frac{4\pi \int_{0}^{r}dx R^{2}(t,x)R'(t,x)\delta(t,x)}{4\pi  \int_{0}^{r}dx R^{2}(t,x)R'(t,x)}
\end{equation}
with the spatial integration in the perturbed metric
and the Friedmann equation (\ref{eq:Friedmann_eq}) is used.

\subsubsection{General time slicing}
Next we will see the situation in the general time slicing with 
the conformally flat spatial coordinates, for which the line element is
given in the form of Eq.~(\ref{eq:metric_spherical_SS}).
Then, $R$ is given by 
\begin{equation}
 R=\psi^{2}ar,
\end{equation}
and the Kodama mass is given by 
\begin{equation}
 M (t,r)=4\pi \int_{0}^{r}dx x^{2}a^{3}\alpha \psi^{6}T^{t}_{\mu}K^{\mu}
\end{equation}
from Eq.~(\ref{eq:Kodama_mass}).
Since $K^{\mu}$ are given by~\cite{footnote1}
\begin{eqnarray}
 K^{t}=-\frac{1}{\alpha\psi^{2}}(\psi^{2}r )',\quad 
 K^{r}=\frac{r}{\alpha\psi^{2}a}(\psi^{2}a)_{,t},\quad
 K^{\theta}=K^{\phi}=0, 
\end{eqnarray}
$M$ is expressed as 
\begin{equation}
M(t,r)=4\pi
 a^{3}\int_{0}^{r}dx  x ^{2}\psi^{4}\left\{-\left[(\rho+p)u^{t}u_{t}+p\right](\psi^{2}x)'+(\rho+p)u^{t}u_{r}\frac{x}{a}(\psi^{2}a)_{,t}\right\}.
\end{equation}
In the flat FRW spacetime, this reduces to 
\begin{equation}
 M_{F}(t,r)=4\pi a^{3}\rho_{b} \int^{r}_{0}dx x ^{2}.
\end{equation} 
We have defined the mass excess $\delta M(t,r)$ as 
the difference 
between the Kodama masses enclosed 
in two spheres with the {\it same areal radius}
in
the perturbed spacetime and the FRW spacetime, i.e., 
\begin{equation}
 \delta M(t,r):=M(t,r)-M_{F}(t,\psi^{2}r).
\end{equation}
Assuming the long-wavelength solutions,
we find 
\begin{equation}
 M(t,r)=4\pi a^{3}\rho_{b}\int_{0}^{r}dx x^{2}(1+\delta) \psi^{6}\left(1+\frac{2r }{\psi}\psi'\right)+O(\epsilon^{3}),
\end{equation}
while the integral in $M_{F}(t,\psi^{2}r)$ can be transformed as follows:
\begin{equation}
 \int_{0}^{\psi^{2}(t,r)r }dx x^{2}=\int_{0}^{\varphi_{t}(r)}dx x^{2}
=\int_{0}^{r}dy \frac{d\varphi_{t}}{dy}\varphi_{t}^{2}
=\int_{0}^{r}dy  y^{2}\psi^{6}\left(1+\frac{2y}{\psi}\psi'\right),
\end{equation}
where $\varphi_{t}(r):=\psi^{2}(t,r) r $. 
Therefore, we find 
\begin{equation}
 \delta M=4\pi a^{3}\rho_{b}\int_{0}^{r}dx  x ^{2}
  \psi^{6}\left(1+\frac{2x }{\psi}\psi'\right)\delta+O(\epsilon^{3}).
\label{eq:SS99_4.28}
\end{equation}
Thus, we have established the theorem that the 
density perturbation integrated over the perturbed spacetime 
coincides with the mass excess to
$O(\epsilon^{2})$ in any time slicing 
which is compatible with the long-wavelength scheme.
The compaction function satisfies
\begin{equation}
 {\cal C}=\frac{1}{2}\bar{\delta}(H_{b}R)^{2}+O(\epsilon^{3}).
\label{eq:compaction_function_general}
\end{equation}
We can confirm that in the comoving slicing, where $u_{r}=0$, 
Eqs.~(\ref{eq:SS99_4.28}) and (\ref{eq:compaction_function_general})
hold without the term of 
$O(\epsilon^{3})$ on the right-hand side.

\section{Spherically symmetric cosmological perturbations}

\subsection{Long-wavelength solutions}
Now that we have the long-wavelength solutions without 
symmetry in Sec.~\ref{sec:long_wavelength_solutions}, it is straightforward to apply those to spherical symmetry.
For convenience, 
we use the spherical coordinates in which the flat metric $\eta_{ij}$ 
takes the following form:
\begin{equation}
 \eta_{ij}dx^{i}dx^{j}=dr^{2}+r^{2}(d\theta^{2}+\sin^{2}\theta d\phi^{2}).
\end{equation}

In the CMC slicing, the solution is given by
Eqs.~(\ref{eq:A_tilde_ij_solution}),
(\ref{eq:h_ij_solution}), (\ref{eq:delta_CMC})--(\ref{eq:xi_CMC}), 
where the expressions for $\xi $, $v_{j}$ and $h_{ij}$ hold
only for the normal coordinates and 
\begin{eqnarray}
 \Psi&=&\Psi(r), 
\label{eq:Psi_r}\\
f&=&-\frac{4}{3}\frac{1}{\Psi^{5}}\frac{1}{r^{2}}\frac{d}{dr}\left(r^{2}\frac{d\Psi}{dr}\right), 
\label{eq:f_Psi}\\
 p_{ij}&=&\frac{1}{\Psi^{4}}
\left[
-\frac{2}{\Psi}\left(\bar{\cal D}_{i}\bar{\cal
		D}_{j}\Psi-\frac{1}{3}\eta_{ij}\bar{\Delta}\Psi\right)+\frac{6}{\Psi^{2}}\left(\bar{\cal
		D}_{i}\Psi\bar{\cal
		D}_{j}\Psi-\frac{1}{3}\eta_{ij}\bar{\cal
		D}^{k}\Psi\bar{\cal D}_{k}\Psi\right)\right].
\label{eq:p_ij_Psi}
\end{eqnarray}

In the comoving slicing, the solution is given by
Eqs.~(\ref{eq:A_tilde_ij_solution}),
(\ref{eq:h_ij_solution}), (\ref{eq:delta_CC})--(\ref{eq:xi_CC}), 
where the expressions for $\xi $, $v_{j}$ and $h_{ij}$ are valid in the
normal coordinates and $f$ and $p_{ij}$ are given by
Eqs.~(\ref{eq:Psi_r})--(\ref{eq:p_ij_Psi}).

\subsection{Asymptotic quasihomogeneous solutions \label{subsec:asymptotic_quasi-homogeneous_solutions}}
Polnarev and Musco~\cite{Polnarev:2006aa} presented asymptotic quasihomogeneous solutions
as cosmological nonlinear perturbations 
based on the Misner-Sharp formulation. We briefly review these solutions
below.

We choose the flat FRW spacetime as the reference spacetime, which 
is given by Eq.~(\ref{eq:spherically_symmetric_diagonal_metric}) with
\begin{eqnarray*}
\alpha=1,~ 
b=a(t),~
R=a(t)r=R_{b}(t,r),~
\rho=\rho _{b}(t),~
M=M_{b}:=\frac{4\pi \rho _{b}R_{b}^{3}}{3},~
U=\dot{a}r=H_{b}R_{b}, ~
H_{b}=\frac{\dot{a}}{a}, 
\end{eqnarray*}
where 
$
U:=\dot{R}/\alpha.
$
The line element is written in the form
\begin{eqnarray*}
ds^{2}=-dt^{2}+a^{2}(t)[dr^{2}+r^{2}(d\theta^{2}+\sin^{2}\theta d\phi^{2})].
\end{eqnarray*}

We define $r_{0}$ as the comoving scale under consideration. Then, 
$R_{0}=a(t)r_{0}$ is the corresponding physical scale.
We define 
\begin{equation}
\varepsilon:=\left(\frac{1}{H_{b}R_{0}}\right)^{2}
=\left(\frac{1}{\dot{a}r_{0}}\right)^{2}
=\left(\frac{1}{H_{b} a r_{0}}\right)^{2}.
\end{equation}
Note that $\varepsilon$ is time dependent and 
\begin{equation}
 \frac{\dot{\varepsilon}}{\varepsilon}=(3\Gamma-2)H_{b}
\end{equation}
implies that $\varepsilon$ increases in time if $\Gamma>2/3$.
The correspondence between the Shibata-Sasaki $\epsilon$ and
Polnarev-Musco $\varepsilon$ is thus given by 
$
 \epsilon^{2}\sim\varepsilon$, 
although the perturbation scheme is somewhat different; Polnarev and
Musco~\cite{Polnarev:2006aa} introduce $\varepsilon$ as a small time-dependent function, while 
Shibata and Sasaki~\cite{Shibata:1999zs} introduce $\epsilon$ as a constant order parameter 
which controls the order of spatial gradients at the fixed time
and which is finally taken to be unity after the expansion.

As for nonlinear cosmological fluctuations, 
we assume that the metric approaches
\begin{equation}
 ds^{2}=-dt^{2}+a^{2}(t)\left[\frac{dr^{2}}{1-K(r)r^{2}}+r^{2}(d\theta^{2}+\sin^{2}\theta d\phi^{2})\right]
\end{equation}
in the limit $\varepsilon \to 0$ and expand $\alpha$, $b$, $R$, $\rho $, $U$ and $M$ as 
\begin{eqnarray*}
&&\alpha=1+\varepsilon \tilde{\alpha}, 
b=\frac{R'}{\sqrt{1-K(r)r^{2}}}(1+\varepsilon
 \tilde{b}), 
R=R_{b}(1+\varepsilon \tilde{R}), \\
&&\rho =\rho _{b}(1+\varepsilon \tilde{\rho}), 
U=H_{b}R(1+\varepsilon \tilde{U}), 
 M=\frac{4\pi}{3}\rho _{b}R^{3}(1+\varepsilon \tilde{M}), 
\end{eqnarray*}
noting that $K(r)=O(\epsilon^{0})$. 
We expand the Einstein equations (\ref{eq:M'})--(\ref{eq:definition_M}) 
in a power series of $\varepsilon$. 
We can find that the first-order functions can be written in terms of
$K(r)$ through Eq.~(\ref{eq:definition_M}).
The concrete expressions are as follows:
\begin{eqnarray}
 \tilde{\alpha}&=&-\frac{3(\Gamma-1)}{3\Gamma+2}\frac{(r^{3}K(r))'}{3r^{2}}r_{0}^{2}+O(\varepsilon), 
\label{eq:tilde_alpha}\\
 \tilde{b}&=&\frac{3(\Gamma-1)}{(3\Gamma-2)(3\Gamma+2)}r\left[\frac{(r^{3}K(r))'}{3r^{2}}\right]'r_{0}^{2}+O(\varepsilon), 
\label{eq:tilde_b}\\
 \tilde{R}&=&-\frac{1}{(3\Gamma-2)(3\Gamma+2)}
\left[(\Gamma-1)\frac{(r^{3}K(r))'}{r^{2}}+K(r)\right]r_{0}^{2}+O(\varepsilon), 
\label{eq:tilde_R}\\
 \tilde{\rho}&=&\frac{3\Gamma}{3\Gamma+2}\frac{(r^{3}K(r))'}{3r^{2}}r_{0}^{2}+O(\varepsilon), 
\label{eq:tilde_rho}\\
 \tilde{M}&=&\frac{3\Gamma}{3\Gamma+2}K(r)r_{0}^{2}+O(\varepsilon), \\
\label{eq:tilde_M}
 \tilde{U}&=&-\frac{1}{3\Gamma+2}K(r)r_{0}^{2}+O(\varepsilon),
\end{eqnarray}
where we have dropped a decaying mode. 

The density perturbation $\bar{\delta} (t,r)$ averaged 
over the region inside the sphere of the radius $r$
is calculated as
\begin{equation}
 \bar{\delta}(t,r)
=\varepsilon \frac{3\Gamma}{3\Gamma+2} K(r)r_{0}^{2}+O(\varepsilon^{2}).
\end{equation}
Expanding $\bar{\delta}(t,r)$ as 
$
 \bar{\delta}(t,r)=\varepsilon \tilde{\bar{\delta}}(t,r), 
$
we find
\begin{equation}
 \tilde{\bar{\delta}}(t,r)=\frac{3\Gamma}{3\Gamma+2}K(r)r_{0}^{2}+O(\varepsilon),
\label{eq:tilde_bar_delta_K_r02}
\end{equation}
which is time independent in the limit $\varepsilon\to 0$. 
This coincides with the mass perturbation $\tilde{M}(t,r)$ 
to $O(\varepsilon^{0})$, i.e.,
$
 \tilde{\bar{\delta}}=\tilde{M}+O(\varepsilon).
$
This is consistent with the full-order 
equation~(\ref{eq:C_bar_delta_comoving}) noting 
$M_{b}=H_{b}^{2}R_{b}^{3}/2$ and $R_{b}=ar$.

For variable equations of state and for more general modes,
see Polnarev and Musco~\cite{Polnarev:2006aa}.
For higher-order solutions with 
respect to $\varepsilon$, 
see Polnarev {\it et al.}~\cite{Polnarev:2012bi}.

\subsection{Equivalence of the two solutions}

As we have seen in Sec.~\ref{subsec:equivalence_slicings}, 
the CMC slicing and comoving slicing
coincide with each other to $O(\epsilon)$.
This means that the spatial metrics $d\Sigma^{2}$
in these slicings are identical to $O(\epsilon)$. 
In the conformally flat coordinates, the spatial metric 
is given by 
\begin{eqnarray*}
d{\Sigma}^{2}=a^{2}(t)\Psi^{4} (\varpi)
\left[d\varpi^{2}+\varpi^{2}(d\theta^{2}+\sin^{2}\theta
 d\phi^{2})\right]
+O(\epsilon^{2}),
\end{eqnarray*}
while in the areal radial coordinates, it is given by 
\begin{equation*}
 d{\Sigma}^{2}=
a^{2}(t)\left[\frac{dr^{2}}{1-K(r)r^{2}}+r^{2}(d\theta^{2}+\sin^{2}
\theta d\phi^{2})\right]+O(\epsilon^{2}),
\end{equation*}
where and hereafter we denote the radial coordinate in
the conformally flat spatial coordinates with $\varpi$
to distinguish between the two coordinates.
The former and the latter are adopted by Shibata and 
Sasaki~\cite{Shibata:1999zs} and Polnarev and Musco~\cite{Polnarev:2006aa}, respectively.
The coincidence of the two metrics implies the following relations:
\begin{eqnarray}
 \begin{cases}
  \Psi(\varpi)^{2}d\varpi=\displaystyle\frac{dr}{\sqrt{1-K(r)r^{2}}}, \\
  \Psi(\varpi)^{2}\varpi=r.
\label{eq:varpi_r}
 \end{cases}
\label{eq:psi_K}
\end{eqnarray}
The above expression can be inverted for $r$ and $K(r)$
in terms of $\varpi$ and $\Psi(\varpi)$ as 
\begin{eqnarray}
\begin{cases}
r=\Psi^{2}(\varpi)\varpi, \\
K(r)r^{2}=1-\left(1+2\displaystyle\frac{\varpi}{\Psi(\varpi)} \frac{d\Psi(\varpi)}{d\varpi}\right)^{2}. 
\end{cases}
\label{eq:psi_to_K}
\end{eqnarray}

Using Eqs.~(\ref{eq:psi_K}) and (\ref{eq:psi_to_K}), we can prove the
following relation:
\begin{equation}
\frac{1}{3}\frac{1}{r^{2}}\frac{d(r^{3}K(r))}{dr}=-\frac{4}{3}\frac{1}{\varpi^{2}\Psi^{5}}\frac{d}{d\varpi}\left(\varpi^{2}\frac{d\Psi}{d\varpi}\right).
\label{eq:equivalence_relation_2}
\end{equation}
Let us prove this identity.
From Eq.~(\ref{eq:psi_to_K}), we find 
\[
 \sqrt{1-K(r)r^{2}}=1+2\varpi \frac{1}{\Psi}\frac{d\Psi}{d \varpi}.
\]
Together with Eq.~(\ref{eq:psi_K}), we find
\[
 \frac{1}{3r^{2}}\frac{d}{dr}=\frac{1}{3\varpi^{2}\Psi^{5}\left(\Psi+2\varpi\frac{d\Psi}{d\varpi}\right)}\frac{d}{d\varpi}.
\]
On the other hand, we find
\begin{equation}
 r^{3}K(r)=-4\varpi^{2}\frac{d\Psi}{d\varpi}\left(\Psi+\varpi\displaystyle\frac{d\Psi}{d\varpi}\right).
\label{eq:Kr^3_varpi_Psi}
\end{equation}
Differentiating the right-hand side with respect to $\varpi$, we find
\begin{eqnarray*}
\frac{d}{d\varpi}\left[\varpi^{2}\frac{d\Psi}{d\varpi}\left(\Psi+\varpi\frac{d\Psi}{d\varpi}\right)\right]=\left(\Psi+2\varpi
		  \frac{d\Psi}{d\varpi}\right)
\frac{d}{d\varpi}\left(\varpi^{2}\frac{d\Psi}{d\varpi}\right)
.
\end{eqnarray*}
From the above equations, we finally obtain Eq.~(\ref{eq:equivalence_relation_2}).

From Eqs.~(\ref{eq:f_Psi}) and (\ref{eq:equivalence_relation_2}), we
find a useful relation
\begin{equation}
 f=\frac{1}{3r^{2}}\frac{d}{dr}(r^{3}K(r)).
\label{eq:f_K}
\end{equation}
Equations~(\ref{eq:psi_K}) 
can be integrated to give $\varpi$ and $\Psi(\varpi)$
in terms of $r$ and $K(r)$ as 
\begin{eqnarray}
 \begin{cases}
  \varpi=r\exp\left[\displaystyle\int^{r}_{\infty}\frac{dx}{x}
\left(\frac{1}{\sqrt{1-K(x)x^{2}}}-1\right)\right], \\
  \Psi(\varpi)=\exp\left[-\displaystyle\frac{1}{2}\displaystyle\int^{r}_{\infty}\frac{dx}{x}\left(\frac{1}{\sqrt{1-K(x)x^{2}}}-1\right)\right],
 \end{cases}
\label{eq:K_to_psi}
\end{eqnarray}
where for simplicity we have assumed 
\begin{eqnarray}
 \lim_{r\to \infty}K(r)r^{2}=0, \quad  \lim_{\varpi\to \infty}\Psi(\varpi)=1.
\end{eqnarray}
In spite of the above equivalence, it 
should be noted that the turning point where $K(r)r^{2}=1$, 
which is a coordinate singularity
in the areal radial coordinates, can be 
overcome in the conformally flat coordinates as
Kopp {\it et al.}~\cite{Kopp:2010sh} point out.

The correspondence between the
Polnarev-Musco variables and Shibata-Sasaki
variables is given by 
\begin{eqnarray}
 \delta&=&\varepsilon \tilde{\rho}+O(\varepsilon^{2}), \\
 \chi&=&\varepsilon\tilde{\alpha}+O(\varepsilon^{2}), \\
 \kappa&=&-\varepsilon\frac{1}{3}[3\tilde{\alpha}-(3\Gamma-2)(3\tilde{R}+r\tilde{R}'+\tilde{b})]+O(\varepsilon^{2}).
\end{eqnarray}
Using the relation (\ref{eq:f_K}), 
we can see that the Polnarev-Musco asymptotic quasihomogeneous
solutions given by Eqs.~(\ref{eq:tilde_alpha})--(\ref{eq:tilde_rho}) 
are equivalent to the long-wavelength solutions 
in the comoving slicing 
given by Eqs.~(\ref{eq:delta_CC})--(\ref{eq:alpha2_CC}).

\subsection{Correspondence relation between the two solutions \label{subsec:correspondence_relation}}

The difference between the Shibata-Sasaki and 
Polnarev-Musco formulations is in the choice of time slicing
and spatial coordinates. In 
Eq.~(\ref{eq:density_perturbation_four_slicings}), 
we establish the correspondence relation
\begin{equation}
 \delta_{C}=\frac{3\Gamma}{3\Gamma+2}\delta_{\rm CMC}+O(\epsilon^{4}).
\label{eq:delta_C_delta_CMC}
\end{equation}
The prefactor on the right-hand side is due to the difference in time
slicing. 
Note that Eq.~(\ref{eq:delta_C_delta_CMC}) implies
\begin{equation}
 \tilde{\bar{\delta}}_{C}=\frac{3\Gamma}{3\Gamma+2}\tilde{\bar{\delta}}_{\rm CMC}+O(\epsilon^{4}),
\label{eq:tilde_bar_delta_C_tilde_bar_delta_CMC}
\end{equation}
where $\bar{\delta}_{\rm CMC}=\varepsilon
\tilde{\bar{\delta}}_{\rm CMC}$.
In particular, we find the useful relation from
Eqs.~(\ref{eq:tilde_bar_delta_K_r02}) and (\ref{eq:tilde_bar_delta_C_tilde_bar_delta_CMC}),
\begin{equation}
 \tilde{\bar{\delta}}_{\rm CMC}=K(r)r_{0}^{2}+O(\epsilon^{2}).
\label{eq:bar_delta_CMC_K}
\end{equation}

It is also interesting to derive the expressions for physical quantities
in terms of both the Shibata-Sasaki variables and Polnarev-Musco variables.
Shibata and Sasaki~\cite{Shibata:1999zs} 
measure the amplitude of the perturbation 
by $\Psi(0)$. This can be expressed in terms of the 
Polnarev-Musco variables as follows:
\begin{eqnarray}
 \Psi(0)=\exp\left[\frac{1}{2}\int^{\infty}_{0}\frac{dr}{r}
\left(\frac{1}{\sqrt{1-K(r)r^{2}}}-1\right)\right].
\end{eqnarray}
Since
\begin{eqnarray}
 \delta M &=&4\pi a^{3}\rho_{b}\int_{0}^{\varpi }dx  x^{
  2}  \Psi^{6}
  \left(1+\frac{2x}{\Psi}\frac{d\Psi}{dx}\right)\delta+O(\epsilon^{3}),
\label{eq:deltaM_delta}
\end{eqnarray}
in both slicings, Eq.~(\ref{eq:delta_C_delta_CMC}) implies
\begin{equation}
 \delta M_{C}=\frac{3\Gamma}{3\Gamma+2}\delta M_{\rm CMC}+O(\epsilon^{3}).
\label{eq:delta_M_C_delta_M_CMC}
\end{equation}
For the CMC slicing, substituting Eqs.~(\ref{eq:delta_CMC})
and (\ref{eq:f_Psi}) into Eq.~(\ref{eq:deltaM_delta}), 
we find 
\begin{eqnarray}
 \delta M_{\rm CMC}
&=&\frac{4\pi a^{3}\rho_{b}}{3a^{2}H_{b}^{2}}\int_{0}^{\varpi}dx (-4)
x \left(2\Psi'+x \Psi''\right)\left(\Psi+2x \Psi'\right)+O(\epsilon^{3})
\nonumber \\
 &=&-2a\varpi^{2}\Psi'\left(\Psi+\varpi\Psi'\right)+O(\epsilon^{3}).
\end{eqnarray}
Therefore, the mass excess uniformly increases in proportion to 
the scale factor.
The compaction function ${\cal C}$ is then calculated as follows:
\begin{equation}
 {\cal C}_{\rm CMC}=\frac{\delta M_{\rm CMC}}{\varpi
  \Psi^{2}a}=\frac{1}{2}\left[1-\left(1+2\frac{\varpi}{\Psi}\frac{d\Psi}{d\varpi}\right)^{2}\right]+O(\epsilon^{3}).
\label{eq:C_CMC}
\end{equation}
This is time independent to
$O(\epsilon^{2})$.
From Eq.~(\ref{eq:delta_M_C_delta_M_CMC}), we have
\begin{equation}
 {\cal C}_{C}=\frac{3\Gamma}{3\Gamma+2}{\cal C}_{\rm CMC}+O(\epsilon^{3}).
\end{equation}
From Eqs.~(\ref{eq:psi_to_K}) and (\ref{eq:C_CMC}), 
we find a very simple relation
\begin{equation}
 {\cal C}_{\rm CMC}=\frac{1}{2}K(r)r^{2}+O(\epsilon^{3}).
\label{eq:C_CMC_K}
\end{equation}
Hence, from Eqs.~(\ref{eq:tilde_bar_delta_K_r02}) 
and (\ref{eq:tilde_bar_delta_C_tilde_bar_delta_CMC}), 
the relation between 
$\tilde{\bar{\delta}}$ 
and the compaction function ${\cal C}$ is given by 
\begin{equation}
 {\cal C}=\frac{1}{2}\tilde{\bar{\delta}}(r)\left(\frac{r}{r_{0}}\right)^{2}+O(\epsilon^{3})
\label{eq:C_K}
\end{equation}
and, in particular, 
\begin{equation}
 {\cal
  C}(t,r_{0})=\frac{1}{2}\tilde{\bar{\delta}}(r_{0})+O(\epsilon^{3})
\label{eq:C_r0_delta_r0}
\end{equation}
in any slicing. In fact, we can see that 
Eqs.~(\ref{eq:C_K}) and (\ref{eq:C_r0_delta_r0})
directly follow from
Eqs.~(\ref{eq:C_bar_delta_comoving}) and 
(\ref{eq:compaction_function_general}).

Before moving onto the comparison of the numerical results, 
we show how the asymptotic condition restricts the functions
$\Psi(\varpi)$ and $K(r)$. If we assume $\Psi(\varpi)\to 1$ and 
$K(r)r^{2}\to 0$ in the asymptotic region, 
from Eq.~(\ref{eq:C_CMC}),
the asymptotic flat FRW condition implies 
$
 \Psi-1=O\left({1}/{\varpi}\right),
$
while the compensation condition implies faster falloff.
If we assume
$
\Psi=1+C_{\psi}/(2\varpi)+O\left({1}/{\varpi^{2}}\right), 
$
we find 
\begin{eqnarray}
 \lim_{r\to \infty}\delta M_{\rm CMC}=aC_{\psi}, \quad 
 {\cal C}_{\rm CMC}\approx \frac{C_{\psi}}{\varpi}.
\end{eqnarray}
The compensated model corresponds to $C_{\psi}=0$. 
In terms of $K(r)$, we find 
the asymptotic flat FRW condition implies 
$
 K(r)= O\left({1}/{r^{3}}\right)
$,
while the compensation condition implies faster falloff.
If we assume 
$
 K(r)=2C_{K}/r^{3}+O\left({1}/{r^{4}}\right)
$,
we find 
\begin{eqnarray}
 \lim_{r\to \infty}\delta M_{\rm CMC}=aC_{K}, \quad 
 {\cal C}_{\rm CMC}\approx \frac{C_{K}}{r}.
\end{eqnarray}
The compensated model corresponds to $C_{K}=0$.

\section{Numerical results}

\subsection{Setup of numerical simulations}

In the decelerated expansion, such as in the radiation-dominated 
era, the Hubble horizon expands in terms of the comoving scale.
In this phase, primordial cosmological perturbations 
which were on superhorizon scale
enter within the Hubble scale, which is called horizon entry.
Much before the horizon entry, 
such primordial cosmological perturbations are naturally described 
by the long-wavelength solutions. 
The argument of Jeans instability strongly suggests that 
the density perturbation which collapses to a black hole
must be of order unity at horizon entry 
because the Jeans scale is comparable with the Hubble scale
if $\Gamma-1=O(1)$. 
Since $\tilde{\bar{\delta}}(r_{0})$ approximately gives the 
averaged density perturbation at horizon entry, the discussion in
Sec.~\ref{subsec:correspondence_relation} implies that the compaction
function ${\cal C}$ and the curvature variable $\Psi$ must have been perturbed by 
order of unity even much before the horizon entry.
Thus, it is natural to prepare the nonlinear long-wavelength 
solutions as initial data sets for PBH formation 
at the moment much before the horizon entry. 

As we have seen, there are two major approaches, the one is 
adopted by Shibata and Sasaki~\cite{Shibata:1999zs} and 
the other is adopted by Polnarev and Musco~\cite{Polnarev:2006aa}.
They are different not only in time slicing and spatial coordinates
but also in amplitude measures and in initial 
curvature profiles, which lead to the 
complexity in the comparison of the numerical results 
between these two approaches.
Since we have already discussed the formulations, we will 
focus on the difference in initial curvature profiles.
Shibata and Sasaki~\cite{Shibata:1999zs} determine the initial data 
by performing iteration to solve the Hamiltonian and momentum constraints
with the conformally flat spatial coordinates
so that the density profile is given by 
\begin{eqnarray}
 \psi^{6} \delta_{\rm CMC} =C_{\delta}\left[\exp\left(-\frac{\varpi^{2}}
{\varpi_{0}^{2}}\right)
-\sigma^{-3}\exp\left(-\frac{\varpi^{2}}{\sigma^{2}\varpi_{0}^{2}}\right)\right]\left(\frac{t}{\varpi_{0}}\right)^{2-\frac{4}{3\Gamma}},
\label{eq:initial_Psi_SS}
\end{eqnarray}
with the boundary condition
\begin{equation}
 \psi=1+\frac{C_{\psi}}{2\varpi}+O(\varpi^{-2})
\end{equation}
as $\varpi\to \infty$,
while $\delta_{\rm CMC}$ and $u_{j~{\rm CMC}}$ 
are those for the long-wavelength solutions,
i.e., Eqs.~(\ref{eq:delta_CMC}) and (\ref{eq:ur_CMC}).
$\varpi_{0}$ is fixed and then $\sigma$ and $C_{\delta}$ 
parametrize the profiles. 
Note that the above model is approximately compensated
because the overdensity in the first term and the underdensity in
the second term in the square brackets 
on the right-hand side of Eq.~(\ref{eq:initial_Psi_SS}) 
cancel out if they are integrated
over the whole flat space with $\psi=1$.
However, since the space is not flat or $\psi$ is not identically unity, 
this model is generally uncompensated.
Another complexity comes from the different amplitude measures. 
Shibata and Sasaki~\cite{Shibata:1999zs} adopt $\Psi(0)$ and 
${\cal C}_{\rm CMC, max}$ as amplitude
measures, where ${\cal C}_{\rm CMC, max}$ is 
the maximum value of ${\cal C}_{\rm CMC}$.

Polnarev and Musco~\cite{Polnarev:2012bi} and Musco and 
Miller~\cite{Musco:2012au} 
give Gaussian-type profiles and top-hat-type profiles.
In both cases, the profiles fall off much 
faster than $r^{-3}$ as $r\to \infty$ and, hence, 
their profiles correspond to exactly compensated models.
In the Gaussian type, $K(r)$ is given by 
\begin{equation}
 K(r)=\left(1+\alpha \frac{r^{2}}{2\Delta^{2}}\right)\exp
  \left(-\frac{r^{2}}{2\Delta^{2}}\right),
\label{eq:Gaussian_type_curvature}
\end{equation}
which is parametrized by $\Delta$ and $\alpha$. 
They adopt
$\tilde{\bar{\delta}}_{C,0}:=\tilde{\bar{\delta}}_{C}(r_{0})$
as an amplitude measure, where 
$r_{0}$ is specified as the smallest positive root of $\tilde{\rho}(r)=0$.  
We will abbreviate $\Psi(0)$, ${\cal C}_{\rm CMC, max}$ and
$\tilde{\bar{\delta}}_{C,0}$ as $\psi_{0}$, ${\cal C}_{\rm max}$ and
$\tilde{\delta}$, respectively.

Hereafter we only keep the lowest order terms and 
neglect higher order terms in terms of $\epsilon$
of the long-wavelength solutions as initial data sets 
because this truncation is numerically justified 
if $\epsilon$ is sufficiently small 
in Polnarev and Musco~\cite{Polnarev:2006aa}. 
Equation~(\ref{eq:tilde_rho}) implies that  
$r_{0}$ is a positive root of the equation
$
 (r^{3}K(r))'=0$, 
while Eq.~(\ref{eq:C_CMC_K}) implies that $r_{1}$ is a 
positive root of the equation
$
 (r^{2}K(r))'=0,
$
where $r=r_{1}$ is the radius at which $C_{\rm CMC}$ takes a maximum.
If we assume a top-hat shape for $K(r)$, 
we can see $r_{0}=r_{1}$
and, hence, 
\begin{equation}
 {\cal C}_{\rm max}=\frac{3\Gamma+2}{6\Gamma}\tilde{\delta}.
\end{equation}
In Sec.~\ref{sec:top-hat_model}, we will see the top-hat curvature 
model in more detail.
For an exact Gaussian model, where 
\begin{equation}
 K(r)=\exp\left(-\frac{r^{2}}{2\Delta^{2}}\right),
\end{equation}
we find $r_{0}=\sqrt{3}\Delta$ and $r_{1}=\sqrt{2}\Delta$ and hence
\begin{equation}
 {\cal C}_{\rm max}=\frac{3\Gamma+2}{9\Gamma}\sqrt{e}\tilde{\delta}.
\label{eq:sqrt_e}
\end{equation}
Note that the above relation is justified only for the exact 
Gaussian profile for
$K(r)$.
There is no one-to-one correspondence
between $\psi_{0}$, ${\cal C}_{\rm max}$ and $\tilde{\delta}$
unless we specify the profile.

\subsection{Comparison of the numerical results}

To test the consistency between the results of Shibata and 
Sasaki~\cite{Shibata:1999zs} and the results of 
Polnarev and Musco~\cite{Polnarev:2006aa}
and Musco and Miller~\cite{Musco:2012au}, 
we will reproduce the former simulation with 
the latter formulation. The matter field is assumed to be a radiation
fluid; i.e., $\Gamma=4/3$ in this subsection.

\begin{figure}[htbp]
 \begin{center}
    \includegraphics[width=0.95\textwidth]{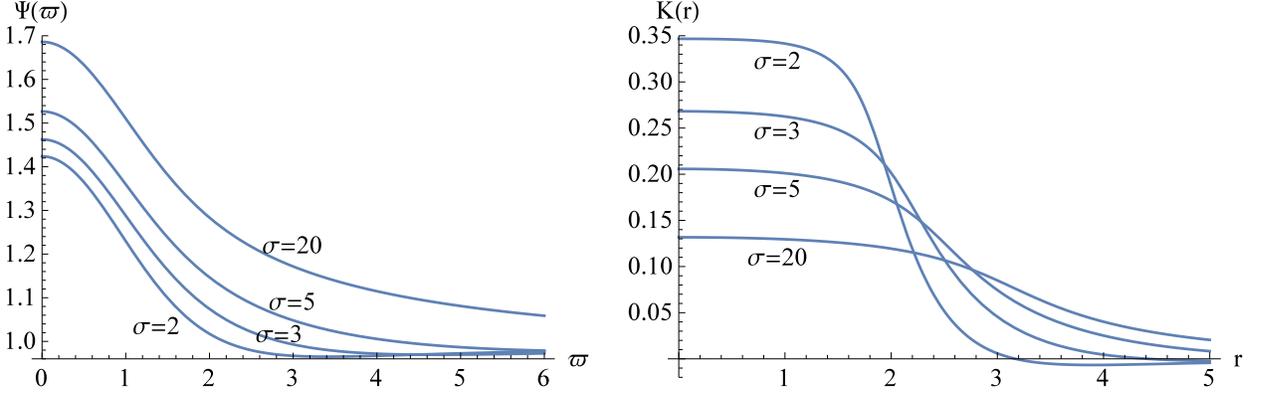} 
\caption{\label{fg:Psi_K} Initial profiles at black hole threshold
 for the model given by Eq.~(\ref{eq:Shibata_Sasaki_initial_data}) 
with the different values of $\sigma$
in terms of $\Psi(\varpi)$ (left panel) and $K(r)$ (right panel). 
}
 \end{center}
\end{figure}

To make discussion clearer, we first integrate 
\begin{equation}
\frac{1}{\varpi^{2}\Psi^{5}}\frac{d}{d\varpi}\left(\varpi^{2}\frac{d\Psi}{d\varpi}\right)
=-2\pi a^{2}\rho_{b}\delta_{\rm CMC},
\end{equation}
which is obtained by Eqs.~(\ref{eq:delta_CMC}) and (\ref{eq:Psi_r}), 
with the source term
\begin{eqnarray}
 \Psi^{6} \delta_{\rm CMC} =C_{\delta}\left[\exp\left(-\frac{\varpi^{2}}
{\varpi_{0}^{2}}\right)
-\sigma^{-3}\exp\left(-\frac{\varpi^{2}}{\sigma^{2}\varpi_{0}^{2}}\right)\right]\left(\frac{t}{\varpi_{0}}\right)^{2-\frac{4}{3\Gamma}}
\label{eq:Shibata_Sasaki_initial_data}
\end{eqnarray}
and the boundary condition 
\begin{equation}
 \Psi=1+\frac{C_{\psi}}{2\varpi}+O(\varpi^{-2})
\end{equation}
as $\varpi\to \infty$. Then, 
we determine $K(r)$ through Eqs.~(\ref{eq:psi_to_K}).
Equivalently, one may define $h(r):=\varpi(r)/r$ and solve 
\begin{equation}
h''+\frac{h^{'2}}{2h}\left(1+\frac{rh'}{h}\right)+\frac{2h'}{r}
=\frac{3C_\delta}{8}h(rh)^{'3}
\left[
\exp(-r^2h^2)-\sigma^{-3}\exp\left(-\frac{r^2h^2}{\sigma^2}\right)
\right],
\end{equation}
with the boundary condition $h=1-C_\psi/r+O(r^{-2})$ as $r\to\infty$. 
It would be useful to note that $h(0)=\Psi(0)^{-2}$. 
Then, $K(r)$ can be calculated by
\begin{equation}
K(r)=\frac{1}{r^2}\left(1-\frac{h(r)^2}{(rh(r))^{'2}}\right).
\end{equation}
The obtained
$K(r)$ generates asymptotic quasihomogeneous solutions
given in Sec.~\ref{subsec:asymptotic_quasi-homogeneous_solutions} 
and we implement numerical simulations based on the Misner-Sharp
formulation adopting those solutions as initial data sets.
Several examples of the initial data sets for both $\Psi(\varpi)$ and $K(r)$
at the black hole threshold 
are plotted in Fig.~\ref{fg:Psi_K} for different values of $\sigma$.
The initial data sets constructed here are identical with 
those in Shibata and Sasaki~\cite{Shibata:1999zs} to $O(\epsilon)$. 

\begin{figure}[htbp]
 \begin{center}
    \includegraphics[width=0.5\textwidth]{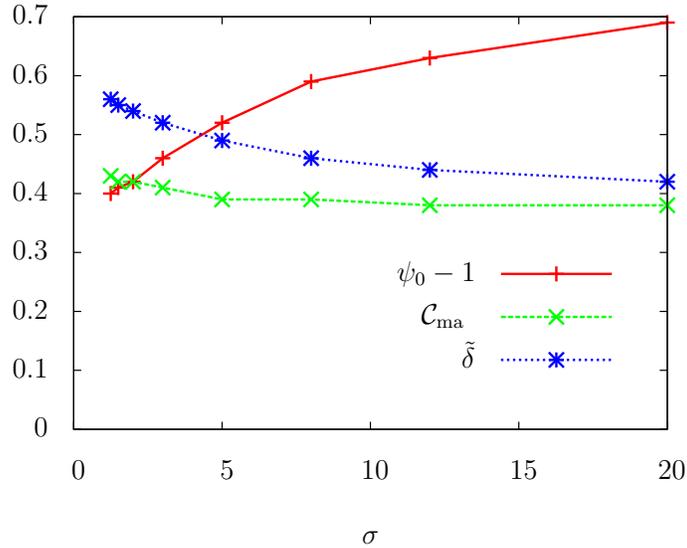} 
\caption{\label{fig:data_nakama} The threshold values of $\psi_{0}$,
${\cal C}_{\rm max}$ and $\tilde{\delta}$ for the model
given by Eq. (\ref{eq:Shibata_Sasaki_initial_data}) for different values
  of $\sigma$.}
 \end{center}
\end{figure}

The details of the numerical code are described in 
Nakama {\it et al.}~\cite{Nakama:2013ica}.
The threshold models for black hole formation for different values of 
$\sigma$ are summarized in Table~\ref{table:threshold_Nakama} and 
Fig.~\ref{fig:data_nakama}, where 
the scales are chosen so that 
$a=a_{f}t^{\frac{2}{3\Gamma}}$, $a_{f}=1$ and $\varpi_{0}=1$.
We can see that the obtained numerical results are fairly 
consistent with Shibata and Sasaki~\cite{Shibata:1999zs}'s 
results by comparing their Figs. 2 and 7 with our Table~\ref{table:threshold_Nakama}.
Note that Shibata and Sasaki~\cite{Shibata:1999zs} estimate 
that the threshold value of 
$\psi_{0}$ is $\simeq 1.79$ in the limit $\sigma\to \infty$.
Here we have also calculated and listed the values for 
$C_{\psi}$ and $\tilde{\delta}$.
We can see that $C_{\psi}$ is negative for these 
threshold models, implying that the models are overcompensated. 
In fact, we have numerically found that $C_{\psi}$ is negative 
also for all nonthreshold models we have calculated
for $1.25\le \sigma\le 20$.
The threshold values of $\tilde{\delta}$ are comparable with 
those obtained by Polnarev and 
Musco~\cite{Polnarev:2006aa} and Musco and 
Miller~\cite{Musco:2012au}. 

\begin{table}[htbp]
\begin{center}
\caption{\label{table:threshold_Nakama}
Black hole formation thresholds for the model given by
 Eq.~(\ref{eq:Shibata_Sasaki_initial_data}) parametrized 
by $\sigma$ and $C_{\delta}$.}
\begin{tabular}{|c|c|c|c|c|c|c|c|c|} \hline
$\sigma$        &  1.25&  1.5&     2&    3&    5&    8&   12&   20\\ \hline\hline
$C_\delta$      &  31.0& 18.9&  13.2& 10.9& 10.4& 10.9& 11.4& 12.1\\ \hline
$C_{\psi}$        & -0.22&-0.24& -0.28&-0.33&-0.41&-0.50&-0.57&-0.63\\ \hline
$\psi_{0}$        &  1.40& 1.41&  1.42& 1.46& 1.52& 1.59& 1.63& 1.69\\ \hline
${\cal C}_{\rm max}$&  0.43& 0.42&  0.42& 0.41& 0.39& 0.39& 0.38& 0.38\\ \hline
$\tilde{\delta} $  &  0.56& 0.55&  0.54& 0.52& 0.49& 0.46& 0.44& 0.42\\ \hline
\end{tabular} 
\end{center}
\end{table}

\begin{figure}[htbp]
 \begin{center}
    \includegraphics[width=0.5\textwidth]{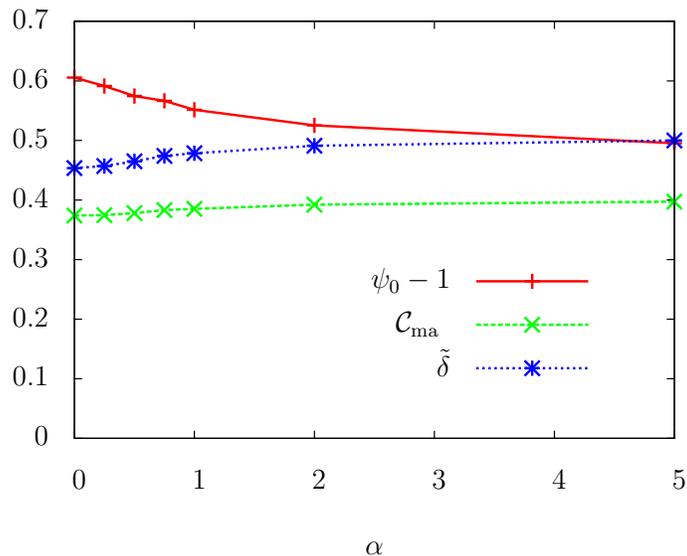} 
\caption{\label{fig:data_koga} The threshold values of $\psi_{0}$,
${\cal C}_{\rm max}$ and $\tilde{\delta}$ for the model
given by Eq.~(\ref{eq:Gaussian_type_curvature}) 
with different values of $\alpha$.}
 \end{center}
\end{figure}

It is also interesting to calculate $\psi_{0}$ and ${\cal C}_{\rm max}$
for the results of Polnarev and 
Musco~\cite{Polnarev:2006aa} and Musco and Miller~\cite{Musco:2012au}.
In fact, this is possible
without reproducing the numerical simulations because the initial 
profiles in Polnarev-Musco and Musco-Miller simulations are given explicitly 
in terms of elementary functions. The results for the Gaussian-type 
profiles are summarized in Table~\ref{table:threshold_Koga} and 
Fig.~\ref{fig:data_koga}.

\begin{table}[htbp]
\begin{center}
\caption{\label{table:threshold_Koga}
Black hole formation thresholds for the model given by
 Eq.~(\ref{eq:Gaussian_type_curvature}) parametrized by 
$\alpha$ and $\Delta$.}
 \begin{tabular}{|c|c|c|c|c|c|c|c|} \hline  $\alpha$ & 0.00 & 0.25 & 0.50 & 0.75 & 1.00 & 2.00 & 5.00 \\ \hline\hline
  $\Delta$ & 1.01 & 0.89 & 0.80 & 0.73 & 0.68 & 0.54 & 0.36 \\ \hline
  $r_0$ & 1.75 & 1.70 & 1.61 & 1.51 & 1.42 & 1.15 & 0.80 \\ \hline
  $r_1$ & 1.43 & 1.40 & 1.35 & 1.29 & 1.22 & 1.01 & 0.71 \\ \hline
  $\psi_{0}$ & 1.61 & 1.59 & 1.57 & 1.57 & 1.55 & 1.53 & 1.50 \\ \hline
  ${\cal C}_{\rm max}$ & 0.37 & 0.37 & 0.38 & 0.38 & 0.39 & 0.39 & 0.40 \\ \hline
  $\tilde{\delta}$ & 0.45 & 0.46 & 0.47 & 0.47 & 0.48 & 0.49 & 0.50 \\ \hline
 \end{tabular}
\end{center}
\end{table}

\subsection{Threshold values of $\tilde{\delta}$, ${\cal C}_{\rm max}$ and $\psi_{0}$}

Hereafter, we denote the threshold values of $\tilde{\delta}$, ${\cal C}_{\rm max}$
and $\psi_{0}$ as $\tilde{\delta}_{c}$, ${\cal C}_{{\rm max}, c}$ and
$\psi_{0, c}$, respectively.
In Table~\ref{table:threshold_Nakama} and Fig.~\ref{fig:data_nakama},
we can see that the larger the $\sigma$
is, the smaller the $\tilde{\delta}_{c}$ becomes.
This tendency is shared by ${\cal C}_{{\rm max}, c}$. 
This can be interpreted as follows. $\varpi_{0}$ gives 
the scale of the overdense region, 
while $\sigma\varpi_{0}$ gives that of the underdense region surrounding
the overdense region and $-\sigma^{-3}$ gives the ratio of the
density perturbation of scale $\sigma\varpi_{0}$ to the that 
of scale $\varpi_{0}$. If $\sigma$ is not so larger than unity, the 
overdensity must be approximately 
compensated by the relatively narrow underdense layer, 
resulting in a sharp transition to the background FRW
universe (see Fig.~\ref{fig:data_nakama}). 
This enhances the effect of the pressure gradient and
suppresses the gravitational collapse, resulting in the larger threshold 
values $\tilde{\delta}_{c}$ and ${\cal C}_{{\rm max}, c}$. 
If $\sigma$ is much larger than unity, the underdense
region spreads to large distances and hence the density there 
is only slightly lower than the FRW spacetime. 
Such a configuration minimizes the effect of pressure gradient force.
This is the reason why 
$\tilde{\delta}_{c}$ and ${\cal C}_{{\rm max}, c}$ 
take minimum values $\simeq 0.42$
and $\simeq 0.38$, respectively, for $\sigma= 20$.

The same tendency can be seen also in the Gaussian-type profiles for 
$K(r)$ as seen in Table~\ref{table:threshold_Koga} and 
Fig.~\ref{fig:data_koga}. 
The larger the $\alpha $ is, the larger the $\tilde{\delta}_{c}$ becomes.
Figure~7 in Musco and Miller~\cite{Musco:2012au}
shows that the larger the $\alpha$ is, the larger the density 
gradient and hence the pressure gradient become.
The minimum value $0.45$ of $\tilde{\delta}_{c}$ is realized for  
the pure Gaussian profile of $K(r)$, i.e., $\alpha=0$. The 
factor $\sqrt{e}/2\simeq 0.82$ in Eq.~(\ref{eq:sqrt_e})
explains the ratio of 
${\cal C}_{{\rm max}, c}\simeq 0.37$ to 
$\tilde{\delta}_{c}\simeq 0.45$. 
As for the sharpest transition model, which is an approximately 
top-hat curvature model, Polnarev and Musco~\cite{Polnarev:2006aa} even get 
$\tilde{\delta}_{c}\simeq 0.66$, which is nearly the 
possible maximum value $2/3$ of 
$\tilde{\delta}$ for the density perturbation~\cite{Harada:2013epa}.

On the other hand, the analytic threshold formula 
obtained in Harada {\it et al.}~\cite{Harada:2013epa}
yields $\tilde{\delta}_{c}\simeq 0.4135$ 
for a radiation fluid. One of the key assumptions
to derive this analytic formula is that the effects of pressure gradient 
force due to the transition from the overdensity to the underdense layer 
can be neglected. From the above consideration, we can say that the 
analytic formula approximately gives the lowest value of 
$\tilde{\delta}_{c}$, which is realized 
if the transition between the 
overdense region and the FRW universe is sufficiently smooth.   

For the equation of state $p=(\Gamma-1)\rho$, the above discussion about 
the threshold of PBH formation in spherical symmetry 
is summarized as follows: 
\begin{equation}
 \tilde{\delta}_{c, {\rm
  min}}<\tilde{\delta}_{c}<\tilde{\delta}_{c, {\rm max}}
\end{equation} 
in the comoving slicing, where 
\begin{eqnarray}
 \tilde{\delta}_{c, {\rm  min}}&\simeq &\frac{3\Gamma}{3\Gamma+2}\sin^{2}\left(\frac{\pi\sqrt{\Gamma-1}}{3\Gamma-2}\right), \\
 \tilde{\delta}_{c, {\rm
  max}}&\simeq &\frac{3\Gamma}{3\Gamma+2}.
\end{eqnarray}
The smoother the transition from the overdensity to the FRW universe is, 
the smaller the $\tilde{\delta}_{c}$ becomes.
The minimum value is realized if the transition is sufficiently smooth,
while the maximum is realized if the transition is sufficiently sharp.

As seen in Tables \ref{table:threshold_Nakama} and \ref{table:threshold_Koga},
it does not seem to affect $\tilde{\delta}_{c}$ and $C_{{\rm max}, c}$
so much  
whether the model is compensated or not.
This can be understood that the dynamics of black hole formation
is determined within the background Hubble length and is not 
affected by a global mass excess. Both the amplitude measures 
$\tilde{\delta}$ and ${\cal C}_{\rm max}$ are quasilocal quantities 
and do not care about perturbations 
on scales much longer than the horizon scale at horizon entry. 

On the other hand, we can see that $\psi_{0,c}$ shows a very
different behavior from $\tilde{\delta}_{c}$ and ${\cal C}_{{\rm max}, c}$.
In Table~\ref{table:threshold_Nakama} and Fig.~\ref{fig:data_nakama}, 
we can see that as $\sigma$
increases and hence the transition is smoother, 
$\psi_{0, c}$ {\it increases}.
The same behavior is also seen in Table~\ref{table:threshold_Koga}
and Fig.~\ref{fig:data_koga}.
As $\alpha$ increases and hence the transition is sharper,
$\psi_{0, c}$ {\it decreases}.
Based on this behavior, Shibata and Sasaki~\cite{Shibata:1999zs}
conclude that if the overdensity 
is surrounded by a low density region, it efficiently collapses.
We have confirmed that 
their conclusion is correct
if one 
directly links the small 
$\psi_{0,c}$ to the efficient production of black holes, which should 
applies if the statistical or probability 
distribution of $\psi_{0}$ is regarded as fixed and 
is centered at its unperturbed value.

To elucidate why neither $\tilde{\delta}_{c}$ nor ${\cal C}_{\rm max}$
but $\psi_{0,c}$ significantly depends on the
behavior of the surrounding region, it is helpful to recall that 
$\Psi$ corresponds to the Newtonian potential for the density 
perturbation in the CMC slicing,
which can be most clearly seen in Eq.~(\ref{eq:SS99_3.5}) with $\kappa=0$.
The Newtonian potential contains the information of the surrounding 
region in contrast to the averaged density and the compaction function.
For example, we can determine the averaged density and the 
compaction function at some radius $r$ from the density distribution
inside the sphere of radius $r$. On the contrary, to determine the 
Newtonian potential and its central value, we need the density 
distribution not only inside but also outside the sphere. 
Since the PBH formation threshold should be determined 
by dynamics inside the Hubble length, it would be most efficiently 
described by quasilocal quantities, such as $\tilde{\delta}_{c}$ and
${\cal C}_{\rm max}$, rather than $\psi_{0,c}$. This is consistent with
the suggestion by Young {\it et al.}~\cite{Young:2014ana}.

At this stage it would be worthwhile to make several comments about the
relationship between the arguments above and Nakama {\it et al.}~\cite{Nakama:2013ica}. 
This paper investigated the PBH formation condition for a much wider
class of initial curvature 
profiles $K(r)$, which is described by 
a function with as many as five parameters, basically using the method
of Polnarev and Musco~\cite{Polnarev:2006aa}.
This function mathematically includes (\ref{eq:Gaussian_type_curvature}) and also 
includes profiles which are very close to a top-hat, as well as profiles which are gentler than a Gaussian. 
In addition, that function was claimed to include the profiles investigated by Shibata and Sasaki, characterized by Eq.~(\ref{eq:initial_Psi_SS}), 
since the function turned out to fit the profiles of Shibata and Sasaki, after being translated into $K(r)$, fairly well with appropriate parameter choice. 
Actually that function was introduced partially in order to express gentler profiles as the ones investigated by Shibata and Sasaki, which are realized when 
$\sigma$ is large. 
However, this statement about the inclusion may seem mathematically inaccurate since 
\cite{Nakama:2013ica} also restricts attention to exactly compensated profiles, 
while Shibata and Sasaki's profiles are strictly speaking overcompensated as has been pointed out above. 
Still, that statement about the inclusion is physically justified since whether 
the initial perturbation is compensated or not does not affect the formation of PBHs that much, 
as has also been pointed out above. 

Though the definition and physical meaning of $\tilde{\delta}$, along with the definition of $r_0$ as the radius of the overdensity, 
is clear and hence it is convenient, 
when one tries to discuss the PBH formation condition for more general profiles more precisely, it is also useful to introduce another phenomenological parameter 
characterizing the amplitude of initial perturbations 
instead of $\tilde{\delta}$, since $\tilde{\delta}$ is too sensitive to the information around $r_0$, which is expected not to affect the formation of PBHs that much. 
In \cite{Nakama:2013ica} a parameter $I$ was phenomenologically introduced, which is similar to $\tilde{\delta}$, as well as $\Delta$, which can be interpreted to measure pressure gradient force. 
It turned out that only these two parameters, characterizing perturbation profiles, are sufficient to describe the formation of PBHs quite well for the generalized class of curvature profiles specified by five parameters. 
In addition, the tendency mentioned above that $\tilde{\delta}_c$ is larger when the transition is 
sharper due to larger pressure gradient force, is also manifest in terms
of these two crucial parameters $I$ and $\Delta$ for the generalized
profiles. 

\section{Analytic results with simplified models}

\subsection{Compensated top-hat density model: Physical interpretation}

Here, we will see why $\psi_{0,c}$ behaves in 
the opposite way to $\tilde{\delta}_{c}$ 
and ${\cal C}_{{\rm max},c}$. 
Since our purpose 
here is to understand this behavior qualitatively, 
we mimic the density profile model (\ref{eq:initial_Psi_SS})
by the following simple function:
\begin{equation}
 \delta_{\rm CMC} =C_{\delta}\left[\Theta(r_{0}-r)
-\sigma^{-3}\Theta(\sigma r_{0}-r)\right]
\left(\frac{t}{r_{0}}\right)^{2-\frac{4}{3\Gamma}},
\label{eq:toy_model}
\end{equation}
where $\Theta(x)$ is Heaviside's step function. 
From Eq.~(\ref{eq:bar_delta_CMC_K}), we find
\begin{eqnarray}
 K(r)=\left\{\begin{array}{cc}
       C'_{\delta}(1-\sigma^{-3})& (0\le r<r_{0})\\
       C'_{\delta}[(1-\sigma^{-3})r_{0}^{3}-\sigma^{-3}(r^{3}-r_{0}^{3})]/r^{3}&(r_{0}\le
	r<\sigma
	 r_{0}) \\
	      0 & (r\ge \sigma r_{0})
	     \end{array}\right. ,
\end{eqnarray}
where we have put
\begin{equation}
 C'_{\delta}:=\frac{4}{9\Gamma^{2}}\frac{a_{f}^{2}}{r_{0}^{2-\frac{4}{3\Gamma}}}C_{\delta}.
\label{eq:C'_delta}
\end{equation}
We can see this model is exactly compensated
and that $r_{0}$ and $r_{1}$ coincide with each other.
For this model, we can find
\begin{eqnarray}
{\cal C}_{\rm max}&=&\frac{1}{2}C'_{\delta}r_{0}^{2}(1-\sigma^{-3}), \\
\tilde{\delta}&=&\frac{3\Gamma}{3\Gamma+2}C'_{\delta}r_{0}^{2}(1-\sigma^{-3}).
\end{eqnarray}
On the other hand, by invoking the linear-order approximation 
in Eqs.~(\ref{eq:delta_f_CMC}) and (\ref{eq:f_Psi}), 
we obtain $\psi_{0}$ as
\begin{equation}
\psi_{0}-1= \frac{3}{8}C'_{\delta}r_{0}^{2}(1-\sigma^{-1}).
\end{equation}
We should note that $\psi_{0}-1$ depends on $\sigma$ in quite 
a different manner from
${\cal C}_{\rm max}$ and $\tilde{\delta}$.

Although this model is very simple, it is still difficult to 
analytically obtain the black hole threshold. The density gradient 
is initially infinite but it becomes finite immediately after 
the time evolution sets in, so that we have to take the 
balance between the gravitational force and the 
pressure gradient force into account in this highly dynamical system.
Here, we make this model more phenomenological noting
the intriguing dependence on $\sigma$.
For this purpose, we introduce three positive constants of order unity,
$c_{1}$, $c_{2}$ and $c_{3}$, 
to simply parametrize the profile dependence as well as
nonlinearity. 
\begin{eqnarray}
\psi_{0}-1&=& c_{1}\frac{3}{8}C'_{\delta}r_{0}^{2}(1-\sigma^{-1}), \\
{\cal C}_{\rm max}&=&c_{2}\frac{1}{2}C'_{\delta}r_{0}^{2}(1-\sigma^{-3}), \\
\tilde{\delta}&=&c_{3}\frac{3\Gamma}{3\Gamma+2}C'_{\delta}r_{0}^{2}(1-\sigma^{-3}),
\end{eqnarray}
These three parameters are unity for the top-hat density
model (\ref{eq:toy_model}).
From the above equations, we can derive the following 
phenomenological relations:
\begin{eqnarray}
 \psi_{0}-1=\frac{3}{4}\frac{c_{1}}{c_{2}}\frac{1-\sigma^{-1}}{1-\sigma^{-3}}
{\cal C}_{\rm
max}=\frac{c_{1}}{c_{3}}\frac{3\Gamma+2}{8\Gamma}\frac{1-\sigma^{-1}}{1-\sigma^{-3}}\tilde{\delta},
\quad 
 {\cal C}_{\rm max}=\frac{c_{2}}{c_{3}}\frac{3\Gamma+2}{6\Gamma}
\tilde{\delta}.
\label{eq:psi0_Cmax_delta_relation_toy}
\end{eqnarray}
Here we just assume that 
the $\sigma$ dependence of 
$\tilde{\delta}_{c}$ and ${\cal C}_{{\rm max}, c}$ 
is very weak. Then, $\psi_{0,c}$ is a monotonically increasing 
function of $\sigma$ for $\sigma>1$.
The above simplistic analysis explains 
the qualitative feature of the numerical results 
shown in Table~\ref{table:threshold_Nakama} and Fig.~\ref{fig:data_nakama}
fairly well. 
In fact, if we choose 
\begin{equation}
 \frac{c_{1}}{c_{2}}\simeq 2.8, \quad \frac{c_{1}}{c_{3}}\simeq 3
\end{equation}
the relations (\ref{eq:psi0_Cmax_delta_relation_toy}) 
agree with 
the numerical results 
in Fig.~\ref{fig:data_nakama} qualitatively,
although the numerical results are obtained 
by the model 
Eq.~(\ref{eq:Shibata_Sasaki_initial_data}) and the Einstein equation is 
solved fully nonlinearly.
The above discussion suggests that the formation criterion 
can be well described by the quasilocal quantities such as
${\cal C}_{\rm max}$ and $\tilde{\delta}$. To translate this 
criterion into the curvature fluctuation $\psi_{0}$,
we need to know the perturbation profile not only in the overdense 
region but also in the surrounding underdense region which may 
spread to large distances.
We should also note that we do not expect that the profile-dependence 
parameters $c_{1}$, $c_{2}$ and $c_{3}$ strongly depend on the equation
of state because these parameters are determined only by the 
initial density profile. 

\subsection{Uncompensated top-hat density model: Environmental effect}

The physical argument about $\psi_{0,c}$ not only explains why neither 
$\tilde{\delta}_{c}$ nor ${\cal C}_{\rm max}$ but $\psi_{0,c}$
is significantly affected by the behavior of the surrounding region
but also suggests that $\psi_{0,c}$ is affected by 
a significant environmental effect 
if the PBH formation results from 
a perturbation on top of a perturbation 
of longer wavelength. Such a situation is expressed in the 
top-hat density model by replacing the compensating underdense region 
with a perturbed region with the density perturbation $\delta_{l}$.
To describe the overdense region of scale $r_{0}$ with the density contrast 
$\delta_{s}$ surrounded by the perturbed region of scale 
$\sigma r_{0}$ with the contrast $\delta_{l}$, we 
generalize the top-hat density model given by 
Eq.~(\ref{eq:toy_model}) as follows:
\begin{equation}
 \delta_{\rm CMC} =C_{\delta}\left[\Theta(r_{0}-r)
+q\Theta(\sigma r_{0}-r)\right]
\left(\frac{t}{r_{0}}\right)^{2-\frac{4}{3\Gamma}},
\label{eq:generalized_toy_model}
\end{equation}
where 
$
 q:={\delta_{l}}/{\delta_{s}}
$.
$K(r)$ is then given by 
\begin{eqnarray}
 K(r)=\left\{\begin{array}{cc}
       C'_{\delta}(1+q)& (0\le r<r_{0})\\
       C'_{\delta}[(1+q)r_{0}^{3}+q(r^{3}-r_{0}^{3})]/r^{3}&(r_{0}\le r<\sigma
	 r_{0}) \\
	      C'_{\delta}(1+q\sigma^{3})r_{0}^{3}/r^{3} & (r\ge \sigma r_{0})
	     \end{array}\right. ,
\end{eqnarray}
and $C'_{\delta}$ is given by Eq.~(\ref{eq:C'_delta}).
So this model has a nonvanishing mass excess and hence is uncompensated
except for $q=-\sigma^{-3}$. 
We can identify $r_{0}$ and $\sigma r_{0}$ with the short wavelength $s$ and 
long wavelength $l$, respectively, so that $\sigma=l/s$. 
The phenomenological 
relations (\ref{eq:psi0_Cmax_delta_relation_toy}) 
should be replaced with 
\begin{eqnarray}
 \psi_{0}-1=\frac{3}{4}\frac{c_{1}}{c_{2}}\frac{1+q \sigma^{2}}{1+q}
{\cal C}_{\rm
max}=\frac{c_{1}}{c_{3}}\frac{3\Gamma+2}{8\Gamma}\frac{1+q
\sigma^{2}}{1+q}\tilde{\delta}
,\quad 
 {\cal C}_{\rm max}=\frac{c_{2}}{c_{3}}
\frac{3\Gamma+2}{6\Gamma}\tilde{\delta},
\label{eq:psi0_Cmax_delta_relation_toy_phenomenology}
\end{eqnarray}
where we have introduced $c_{1}$, $c_{2}$
and $c_{3}$ to simply parametrize the profile dependence and nonlinearity.
For the exact top-hat density model given by Eq.~(\ref{eq:generalized_toy_model}),
we identify all of $c_{1}$, $c_{2}$ and $c_{3}$ with unity.
Note that $\delta_{l}\ge -1$ if we assume the density field is
everywhere nonnegative.

We should here note that to define $\tilde{\delta}$ we invoke 
averaging over the overdense region. However, if $\delta_{l}>0$ in this 
model, the radius of the overdense region is not $r_{0}$ but $\sigma
r_{0}$. This is a subtle issue in the definition of the averaged 
density perturbation. We here continue to take $r_{0}$. 
In fact, if $\sigma$ is much larger than unity, this ambiguity is 
rather in the choice of the averaging length. If we want to 
discuss the formation of black hole in the scale
of $r_{0}$, we should 
make averaging with the scale of $r_{0}$.

Suppose that $q> -1$ and that ${\cal C}_{{\rm max}, c}$ and 
$\tilde{\delta}_{c}$ very weakly depend on the perturbations of 
longer wavelength.
Equation~(\ref{eq:psi0_Cmax_delta_relation_toy_phenomenology}) 
implies that $\psi_{0,c}$ is increased (decreased) when the overdense region of short wavelength scale $r_{0}$
is surrounded by an overdense (underdense) 
region of long-wavelength scale $\sigma r_{0}$ with $\sigma>1$ and
$\delta_{l}>-\delta_{s}$.
The larger the absolute value of the density perturbation ratio $|q|$
or the larger the scale ratio of the perturbations $\sigma$, 
the larger the environmental effect becomes.
On the other hand, the requirement $K(r)r^{2}<1$ at $r=\sigma r_{0}$
implies that 
$q\sigma^{2}\lesssim (1+q)\tilde{\delta}_{c}^{-1}$
for the threshold model
in the case of $\sigma\gg 1$ and hence the
environmental effect in $\psi_{0,c}$ remains 
of order
unity if $q>0$.
On the other hand, if $q<0$, that is, if the overdense region is
located in a larger underdense region,
there seems no limit on the environmental effect, although this 
must be interpreted with caution because the above estimate 
of $\psi_{0}$ only relies on the linear-order analysis.
Hence, 
if one links the small value of $\psi_{0,c}$
to the efficient production of PBHs, one may conclude that   
the PBH production is significantly enhanced.
It should be noted that the above analysis is simplistic 
and further analysis is necessary in this context.
\subsection{Top-hat curvature model: Sharpest transition
  \label{sec:top-hat_model}}

We can get an analytic expression for the top-hat
curvature model, where $K(r)$ is given by 
\begin{eqnarray}
 K(r)=\Theta(\Delta-r),
\end{eqnarray}
with $0<\Delta\le 1$. For $0\le r<\Delta$, we can find 
\begin{equation}
 \int^{r}_{0}\frac{dx}{x}\left(\frac{1}{\sqrt{1-K(x)x^{2}}}-1\right)
=-\ln \frac{1+\sqrt{1-r^{2}}}{2}.
\end{equation}
Then, Eqs.~(\ref{eq:K_to_psi}) yield
\begin{eqnarray}
 \varpi =  r\frac{1+\sqrt{1-\Delta^{2}}}{1+\sqrt{1-r^{2}}}
\quad 
\mbox{or} \quad 
 r=\frac{2
 (1+\sqrt{1-\Delta^{2}})\varpi}{(1+\sqrt{1-\Delta^{2}})^{2}+\varpi^{2}}
\end{eqnarray}
for $0<r<\Delta$ or $0<\varpi<\Delta$, while 
$r=\varpi$ for $r\ge \Delta$ or $\varpi\ge \Delta$,
and 
\begin{eqnarray}
 \Psi^{2}=\frac{1+\sqrt{1-r^{2}}}{1+\sqrt{1-\Delta^{2}}}
\quad 
\mbox{or} 
\quad 
\Psi^{2}
=\frac{2(1+\sqrt{1-\Delta^{2}})}{(1+\sqrt{1-\Delta^{2}})^{2}+\varpi^{2}}
\label{eq:Psi_top-hat}
\end{eqnarray}
for $0<r<\Delta$ or $0<\varpi<\Delta$, 
while $\Psi=1$ for  $r\ge \Delta$ or $\varpi\ge \Delta$.

Equations~(\ref{eq:delta_CC}), (\ref{eq:tilde_bar_delta_K_r02})
and (\ref{eq:C_CMC_K}) give $\delta_{C}$, $\tilde{\bar{\delta}}_{C}$
and ${\cal C}_{\rm CMC}$, where Eq.~(\ref{eq:f_K}) gives
\begin{eqnarray}
f=\Theta(\Delta-r)-\frac{\Delta}{3}\delta(\Delta-r).
\end{eqnarray}
The top-hat curvature model is clearly 
distinct from the top-hat density model because of the delta function
in the density perturbation.
This model is unphysical because 
the density perturbation $\delta$ is infinitely
negative at the transition $r=\Delta$
from the overdense region to the FRW region.
However, if the continuous model has a very sharp transition, 
then we expect that the top-hat curvature model may describe the
dynamics of the continuous model approximately. 

The top-hat curvature model has infinite density gradient and hence
infinite pressure gradient force for general equations of state.
Moreover, since the density perturbation itself is negatively 
infinite at the transition 
due to the presence of the delta function, 
the pressure gradient force always dominates 
gravitational attraction and hence prevent the model from collapsing 
to a black hole except for the case in which the overdense region is 
initially trapped. This implies that the criterion for the black hole formation 
should be given by $K(r)r^{2}=1$ at the transition, i.e., $\Delta=1$,
because the coordinate singularity at which $K(r)r^{2}=1$ implies 
a marginally trapped surface~\cite{Harada:2013epa}.
This means 
$
 \tilde{\delta}_{c}=3\Gamma/(3\Gamma+2)$ and 
${\cal C}_{{\rm max}, c}=1/2$, respectively.
The result of the numerical simulation by Polnarev and Musco~\cite{Polnarev:2006aa} 
for the sharpest top-hat-type profile is 
consistent with this argument. 
If we substitute $\Delta=1$ and $r=\varpi=0$ into 
Eq.~(\ref{eq:Psi_top-hat}), we find $\psi_{0}=\sqrt{2}$. 
This is close to $\psi_{0,c}$ for $\sigma$ not 
so larger than unity as seen in Table \ref{table:threshold_Nakama}.
The similar trend is also seen in Table \ref{table:threshold_Koga}
for the larger values of $\alpha$.
Since these models have a steeper transition from the overdense region 
to the FRW universe, our expectation is supported by the numerical result.
That is, the threshold values for the model with a sufficiently steep
transition will be approximately given by those for the top-hat model,
$\tilde{\delta}_{c}=3\Gamma/(3\Gamma+2)$, ${\cal C}_{{\rm max},c}=1/2$ and
$\psi_{0,c}=\sqrt{2}$.

\subsection{Double top-hat curvature model: Environmental effect}
Next we move onto a double top-hat curvature model, for which $K(r)$ 
is given by 
\begin{equation}
 K(r)=(1-A)\Theta(\Delta_{1}-r)+A\Theta(\Delta_{2}-r),
\label{eq:double_top-hat_curvature_model}
\end{equation} 
where $A$ is constant, $0<\Delta_{1}\le \Delta_{2}$ 
and we require $K(r)r^{2}\le 1$ for 
the avoidance of the coordinate singularity. 
A similar model is used in Nakama~\cite{Nakama:2014fra} in a different context~\cite{footnote2}.
We now 
extract an environmental effect from this simple model. 
For this model, Eqs.~(\ref{eq:delta_CC}), (\ref{eq:tilde_bar_delta_K_r02})
and (\ref{eq:C_CMC_K}) give $\delta_{C}$, $\tilde{\bar{\delta}}_{C}$
and ${\cal C}_{\rm CMC}$, where Eq.~(\ref{eq:f_K}) gives
\begin{eqnarray}
 f=(1-A)\left[\Theta(\Delta_{1}-r)-\frac{\Delta_{1}}{3}\delta(\Delta_{1}-r)\right]+A\left[\Theta(\Delta_{2}-r)-\frac{\Delta_{2}}{3}\delta(\Delta_{2}-r)\right], 
\end{eqnarray}
and from Eq.~(\ref{eq:equivalence_relation_2})
we can find 
\begin{equation}
\psi_{0}=\sqrt{\frac{2(1+\sqrt{1-\Delta_{1}^{2}A})}{(1+\sqrt{1-\Delta_{1}^{2}})
(1+\sqrt{1-\Delta_{2}^{2}A})}}.
\end{equation}
We can see that the single top-hat curvature model is reproduced by putting $A=0$
or $A=1$ or $\Delta_{1}=\Delta_{2}$. We can make identification
$s=\Delta_{1}$, $l=\Delta_{2}$ and $q=\delta_{l}/\delta_{s}=A/(1-A)$,  
and hence the central overdense region is
surrounded by the lower but still overdense region if $0<A<1$, while the
surrounding region is underdense if $A<0$.
We should however note the delta function contributions 
in the density profile again.
  
\begin{figure}[htbp]
 \begin{center}
    \includegraphics[width=0.5\textwidth]{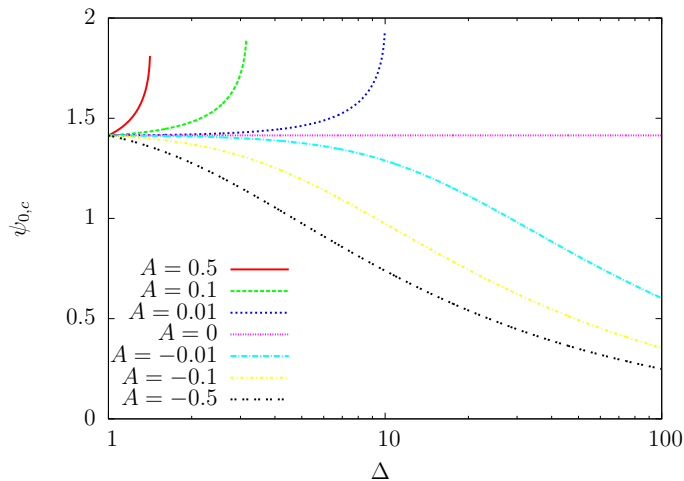}
\caption{\label{fig:double_top-hat_thresholds} 
The threshold values of $\psi_{0,c}$ for the double top-hat curvature
  model given by Eq.~(\ref{eq:double_top-hat_curvature_model}) as
  functions of $\Delta_{2}$ with different values of $A$.}
 \end{center}
\end{figure}

Similarly to the single top-hat curvature model, 
the threshold for black hole formation in this model is given by
$
 \mbox{max}(\Delta_{1}^{2}, \Delta_{2}^{2}A)=1
$.
Let us focus on the cases where $\Delta_{2}^{2}A<1$.
Then, $\Delta_{1}=1$ gives a threshold
and hence $\tilde{\delta}_{c}=3\Gamma/(3\Gamma+2)$ 
and ${\cal C}_{{\rm max},c}=1/2$.
Therefore, $\tilde{\delta}_{c}$ and ${\cal C}_{{\rm max},c}$
do not depend on the surrounding region.
However, this is not the case for 
$\psi_{0,c}$, which is calculated as
\begin{equation}
\psi_{0,c}=\sqrt{\frac{2(1+\sqrt{1-A})}{1+\sqrt{1-\Delta_{2}^{2}A}}}.
\end{equation}
Figure~\ref{fig:double_top-hat_thresholds} shows $\psi_{0,c}$ as a
function of $\Delta_{2}$ for different values of $A$.
Noting $\Delta_{2}>\Delta_{1}=1$ and $\Delta_{2}^{2}A<1$, we can conclude that 
$\psi_{0,c}>\sqrt{2}$ for $A>0$, while $\psi_{0,c}<\sqrt{2}$ for $A<0$.
If $\Delta_{2}(>1)$ is fixed, $\psi_{0,c}$ 
monotonically increases
from $\sqrt{2/\Delta_{2}}$ to $\sqrt{2(1+\sqrt{1-1/\Delta_{2}^{2}})}$
as $A$ increases from $-\infty$ to $1/\Delta_{2}^{2}$. 
If $A\in (0,1)$ and $A$ is fixed, $\psi_{0,c}$
monotonically increases from $\sqrt{2}$ to $\sqrt{2(1+\sqrt{1-A})}$ as
$\Delta_{2}$ increases from $1$ to $1/\sqrt{A}$.
If $A<0$ and $A$ is fixed, $\psi_{0,c}$
monotonically decreases from $\sqrt{2}$ to $0$ as
$\Delta_{2}$ increases from $1$ to $\infty$. 
Thus, the existence of the surrounding overdense (underdense) 
region of longer wavelength increases (decreases) $\psi_{0,c}$
and, hence, suppresses (enhances) the PBH production
if one assumes that 
the smaller $\psi_{0,c}$ is directly related to higher
production rate. Moreover, the environmental effect on 
$\psi_{0,c}$ is bounded if the density perturbation of longer wavelength 
is positive, while $\psi_{0,c}$ even gets smaller than the
unperturbed value if the wavelength $l$ of the underlying negative density
perturbation is sufficiently long and $q=\delta_{l}/\delta_{s}$ is
fixed, and approaches 0 in the limit where $l$ is infinitely long.
The qualitative behavior of $\psi_{0,c}$ 
is common for both the uncompensated top-hat density model 
and the double top-hat curvature model, in spite of the 
physical difference between the models.

\section{Conclusion}

We have constructed cosmological nonlinear perturbation solutions 
with the long-wavelength scheme in the CMC, uniform-density, comoving and geodesic slicings 
without assuming symmetry. 
These solutions are generated by only a master variable 
$\Psi=\Psi(x^{k})$, 
where the perturbations from the flat FRW solution can be 
arbitrarily large.
We have also derived the explicit relation among these four slicings.
One of the interesting applications is to construct initial data
for primordial structure formation with and without 
spherical symmetry. For example,  
we can study nonspherical effects on the 
PBH formation, which are expected to be important especially 
for the soft equation of state.

Then, we have presented two distinct formulations of spherically symmetric 
spacetimes and the definitions of mass excess,
compaction function and averaged density perturbation as spatial gauge
invariant but slicing dependent perturbation quantities.
Based on the general formulation of long-wavelength solutions, 
we have constructed spherically symmetric long-wavelength solutions 
in the CMC slicing and in the comoving slicing.
We have elucidated the relation between the two spatial
coordinates, the conformally flat coordinates and the areal radial 
coordinates. Using these solutions and relations, we have established
the equivalence between the long-wavelength solutions given in 
Shibata and Sasaki~\cite{Shibata:1999zs} and the asymptotic quasihomogeneous solutions 
given in Polnarev and Musco~\cite{Polnarev:2006aa}, 
both of which have been used as initial data sets 
for the simulations of PBH formation.

Using this equivalence, we have reproduced the numerical simulation 
by Shibata and Sasaki~\cite{Shibata:1999zs} with the numerical code based on the formulation 
by Polnarev and Musco~\cite{Polnarev:2006aa} and obtained the
results which agree with the result of 
Shibata and Sasaki~\cite{Shibata:1999zs}.
We have also calculated $\psi_{0,c}$ for the numerical simulation
by Polnarev and Musco~\cite{Polnarev:2006aa}. 
Combining these results, we have discussed that the smoother the
transition from the overdense region to the FRW universe is, the smaller 
the $\tilde{\delta}_{c}$ and ${\cal C}_{{\rm max}, c}$ become. 
The minimum values of $\tilde{\delta}_{c}$ and ${\cal C}_{{\rm max}, c}$ are 
attained if the transition 
from the overdense region to the homogeneous universe is the smoothest.
We have discussed that the analytic threshold formula obtained by
Harada {\it et al.}~\cite{Harada:2013epa} should apply for this case, while the analytic formula for the possible maximum value
can be regarded as the threshold value for a sufficiently sharp transition. 
We have 
found the relation 
among $\psi_{0}$, ${\cal C}_{\rm max}$ and $\tilde{\delta}$ by using 
the compensated top-hat density model and 
interpreted why $\psi_{0,c}$ shows 
an apparently opposite behavior within the (approximately)
compensated models to those 
of $\tilde{\delta}_{c}$ and ${\cal C}_{{\rm max},c }$. 
Moreover, generalizing the top-hat 
density model to the uncompensated one, 
we suggest an environmental 
effect on $\psi_{0,c}$ from the underlying long-wavelength 
perturbations. This is also supported by 
the double top-hat curvature model, which can be treated in a fully 
analytic and nonlinear manner. 
We conclude that even if $\tilde{\delta}_{c}$ and 
${\cal C}_{{\rm max}, c}$ are not sensitive to the 
density profiles in the surrounding region, $\psi_{0,c}$ 
can be significantly 
larger (smaller) if the density perturbation of the wavelength 
in which we are interested is in
the underlying positive (negative) density perturbation of much longer wavelength.

\acknowledgments
The authors are grateful 
to C. Byrnes, B.~J.~Carr, K. Kohri, I. Musco, K.-I. Nakao, K. Nakamura, A. Naruko, M. Sasaki,
M. Siino, T. Kobayashi, J. Yokoyama, S. Yokoyama, 
K. Ogasawara,
T. Shiromizu and J. White for fruitful discussion and useful comments.
T.H. was partially supported by the Grant-in-Aid No. 26400282 for Scientific Research Fund of the
Ministry of Education, Culture, Sports, Science and Technology, Japan.
T.N. was partially supported by Grant-in-Aid for JSPS Fellows No. 25.8199.

\end{document}